\documentclass[reprint,aps,english,twocolumn,notitlepage]{revtex4-2}
\usepackage[utf8]{inputenc}
\usepackage[T1]{fontenc}
\usepackage{babel}
\usepackage{epsfig,epsf,psfrag}
\usepackage{pslatex}
\usepackage{epic,eepic}
\usepackage[colorlinks=true,allcolors=blue]{hyperref}
\usepackage{float}
\usepackage{chngcntr}
\usepackage{color}
\usepackage{graphicx,wrapfig,setspace}
\usepackage[caption=false]{subfig} 
\usepackage{ragged2e} 
\DeclareCaptionJustification{justified}{\justifying}
\usepackage{pstricks}
\usepackage{mathtools}
\usepackage{amssymb}
\usepackage{amsmath}
\usepackage{txfonts}
\usepackage{centernot}
\usepackage{multirow}
\usepackage[normalem]{ulem}
\useunder{\uline}{\ul}{}
\usepackage{natbib}
\usepackage{fancyhdr}
\vspace{-10.0cm}

\begin{document}
\title{Probing the topological character of superconductors via non-local Hanbury-Brown and Twiss correlations}
\author{Tusaradri Mohapatra, Subhajit Pal}
\author{Colin Benjamin}
\email{colin.nano@gmail.com}
\affiliation{School of Physical Sciences, National Institute of Science Education \& Research, HBNI, Jatni-752050, India}
\affiliation{Homi Bhabha National Institute, Training School Complex, Anushaktinagar, Mumbai 400094, India}
\begin{abstract}
Superconductors can be classified as topological or not based on whether time-reversal symmetry (TRS), chiral symmetry, and particle-hole symmetry are preserved or not. Further, topological superconductors can also be classified as chiral or helical. In this paper, using Hanbury-Brown and Twiss (HBT) shot noise correlations and the non-local conductance, we probe metal/2D unconventional superconductor/metal junctions to understand better the pairing topological vs. non-topological or helical vs. chiral or nodal vs. gapful. We see that HBT correlations are asymmetric as a function of bias voltage for non-topological superconductors, whereas they are symmetric for topological superconductors irrespective of the barrier strength. Topological superconductors are associated with Majorana fermions which are important for topological quantum computation. By distinguishing topological superconductors from non-topological superconductors, our study will help search for Majorana fermions, which will aid in designing a topological quantum computer.
\end{abstract}

\maketitle

\section{Introduction}
Conventional spin-singlet $s$-wave superconductors have spherically symmetric order parameters wherein pairing potential is independent of the direction of incident electrons. Any deviation from this is defined as an unconventional superconductor, e.g., $p$-wave or $d$-wave~\cite{UC}. In 2D unconventional superconductors, zero bias conductance peak (ZBCP) may indicate the presence of Majorana zero modes and thus indicate the topological character of superconductors~\cite{ZBCP, ZBCP-TRS}, but does not provide sufficient evidence for the pairing symmetry of the topological superconductor. ZBCP can not distinguish pairing symmetries of different topological superconductors, e.g., chiral-$d$, chiral-$p$, or helical-$p$. Moreover, distinguishing different topological pairings is important as topological superconductors are building blocks of Majorana zero modes, potentially crucial in topological quantum computation~\cite{majorana, topological}.

Among the many proposals to detect pairing symmetry of unconventional superconductors, a well-known method is the Knight shift measurement in a nuclear magnetic resonance (NMR) experiment~\cite{Leggett}. Invariance of the Knight shift to change in temperature below T$_c$ is strongly suggestive of spin-triplet pairing~\cite{Knightshift}, e.g., chiral-$p$, and distinguishes it from the spin-singlet pairing, e.g., chiral-$d$. Now, both chiral-$p$ and chiral-$d$ are topological superconductors. Nevertheless, how do we distinguish chiral-$p$ from chiral-$d$ or helical-$p$? A different method, the current and magnetic field inversion (CFI) symmetry test of time-reversal symmetry (TRS), can be used to discriminate between chiral and helical superconductors, as CFI preserves TRS for helical superconductors. At the same time, it breaks TRS for chiral superconductors~\cite{CFI}. However, these tests are not $100\%$ full proof, so we propose two additional tests, the non-local differential shot noise cross-correlations and non-local HBT correlations in metal/2D unconventional superconductor/metal junction, to distinguish between topological (chiral-$p$, chiral-$d$, helical-$p$, $p_x \hat{x}$ and $p_x \hat{y}$) and non-topological ($s$, $p_x \hat{z}$, $d_{x^2-y^2}$ and $d_{xy}$) based on whether spin rotational symmetry is present or not in 2D. Among these, chiral-$p$, chiral-$d$, helical-$p$, and $s$-wave are gapful, while $p_x \hat{x}$, $p_x \hat{y}$, $p_x \hat{z}$, $d_{xy}$, and $d_{x^2-y^2}$ are nodal. As shown in Ref~\cite{tenfold}, different pairing symmetries of superconductors have been categorized as topological or non-topological via ten-fold classification. {As far as we are aware, shot noise correlations have not yet been used to identify pairing symmetries of nodal triplet superconductors, such as $p_x \hat{x}$, $p_x \hat{y}$ or $p_x \hat{z}$}. Hence, we also include these pairing symmetries in our study with HBT noise to distinguish topological from non-topological pairings and gapful from nodal. To decipher the pairing symmetry of unconventional superconductors, quantum transport in superconducting hybrid junctions via measurement of non-local differential conductance has been helpful~\cite{chiral-p}. However, shot noise or non-local HBT correlations can give more information about the Cooper pair splitting, which may help differentiate between pairing symmetries of 2D unconventional superconductors. We not only focus on cross-correlations but also extend our study to differential shot noise cross-correlations.

The key take-home messages of our study are that HBT or shot noise correlations are asymmetric as a function of bias voltage for non-topological superconductors. In contrast, they are symmetric for topological superconductors irrespective of the barrier strength. Further, we show how different topological pairings like nodal or gapful and chiral or helical can be distinguished from each other via the sign of HBT correlations or their zero bias nature for both transparent and tunneling interfaces. Both differential shot noise and HBT correlations thus serve as effective tools to discriminate between pairing symmetries which we explain in detail in this manuscript.

The manuscript is organized as follows: we give an overview of the possible pairing symmetries of 2D unconventional superconductors in section~\ref{pairing sym}. Section~\ref{2dBTK} deals with the $2D$ BTK approach and how it is used to calculate differential shot noise correlations, shot noise cross-correlations, and non-local conductance in our setup. {Next, in section~\ref{theory} we describe our setup, which is a $2D$ normal metal($N_1$)/insulator(I)/unconventional superconductor(US)/insulator(I)/normal metal ($N_2$) junction for each pairing symmetry.} We then discuss the respective wave functions and boundary conditions necessary to calculate non-local conductance, and shot noise, i.e., HBT correlations and differential shot noise correlations. It is followed by a discussion on the results, first for non-local conductance and differential shot noise cross-correlations and then for HBT cross-correlations. Finally, we provide a discussion on why we see what we see and summarize how different processes contribute to HBT correlations in the subsection "Processes in Play" via Tables~\ref{pplay tun} and \ref{pplay trans}. We finally conclude with a comparison between different pairing symmetries: topological (gapful helical vs. gapful chiral vs. nodal) vs. non-topological using HBT as well as differential shot noise cross-correlations in Tables~\ref{table dsn} and \ref{table sn}. In the Appendix, we have elaborated first on the crossed Andreev conductance and elastic co-tunneling and then in detail on the components of shot noise cross-correlations.

\section{Pairing symmetries}
\label{pairing sym}

\begin{widetext}

\begin{table}
\caption{Pairing symmetries of 2D unconventional superconductors with examples. The examples are from experimental papers where such pairing is the most likely. It is by no means complete confirmation. Herein lies the motivation of our work, which is to provide another method to probe the pairing symmetry of unconventional superconductors unambiguously.}
\scalebox{0.85}{\begin{tabular}{|c|c|c|c|c|c|c|c|c|c|c|}
\hline
\begin{tabular}[c]{@{}c@{}}Topology of 2D\\Superconductors\end{tabular} & Type & Pairing & Parity & \begin{tabular}[c]{@{}c@{}}Symmetry\\classification \cite{tenfold} \end{tabular} & $\textbf{d}(\textbf{k})$ & $\textbf{d}(\theta)$ & $\psi(\textbf{k})$ & $\psi(\theta)$ & $\sigma_c$ vs. $z$ & Examples \\
\hline
\multirow{4}{*}{\begin{tabular}[c]{@{}c@{}}Non\\Topological\end{tabular}} & Gapful & $s$ & Even & CI & 0 & 0 & 1 & 1 & \includegraphics[scale=0.08]{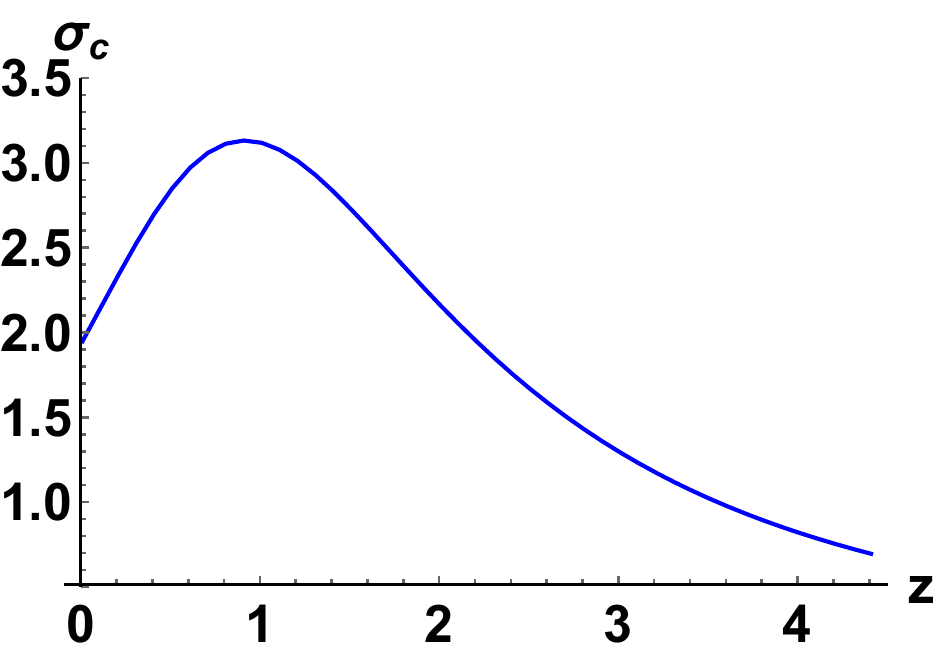} & Al, Pb~\cite{swave-ex} \\
\cline{2-11}
& \multirow{3}{*}{Nodal} & $p_x \hat{z}$ & Odd & AIII & $k_x \hat{z}$ & $\cos{\theta} \hat{z}$ & 0 & 0 & \includegraphics[scale=0.1]{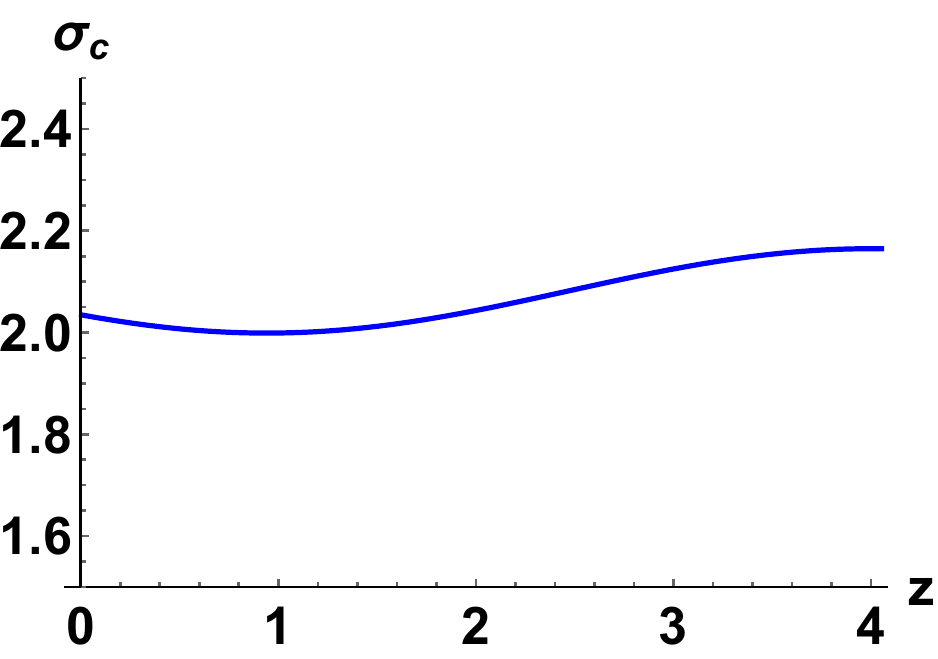} & - \\
\cline{3-11}
& & $d_{x^2-y^2}$ & Even & CI & 0 & 0 & $k^2_x-k^2_y$ & $\cos{2 \theta}$ & \includegraphics[scale=0.1]{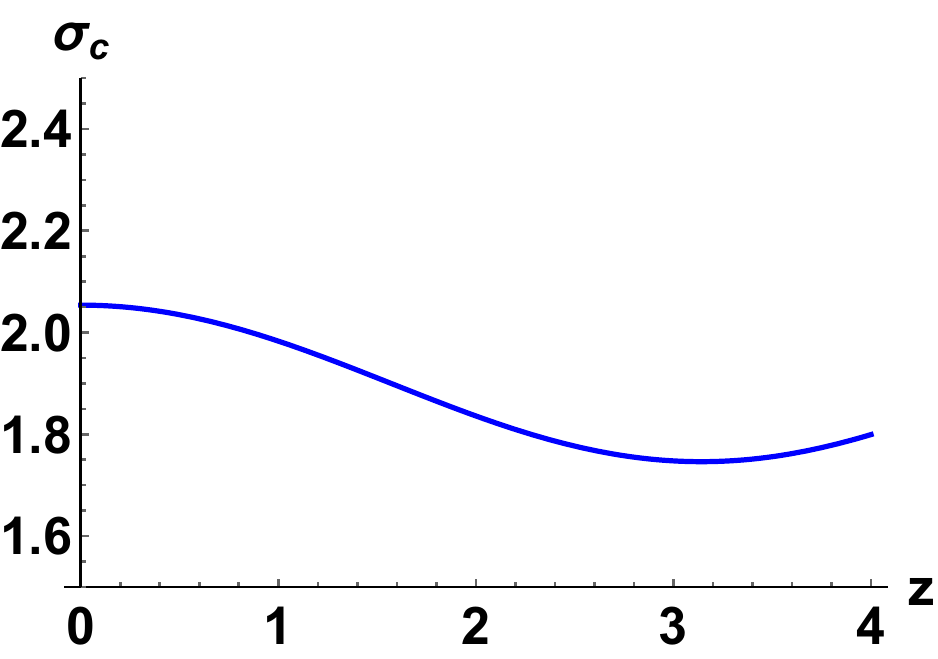} & YBa$_{2}$Cu$_3$O$_{7-\delta}$~\cite{dx2y2ex} \\
\cline{3-11}
& & $d_{xy}$ & Even & CI & 0 & 0 & $ 2 k_x k_y $ & $ 2 \cos{\theta} \sin{\theta}$ & \includegraphics[scale=0.1]{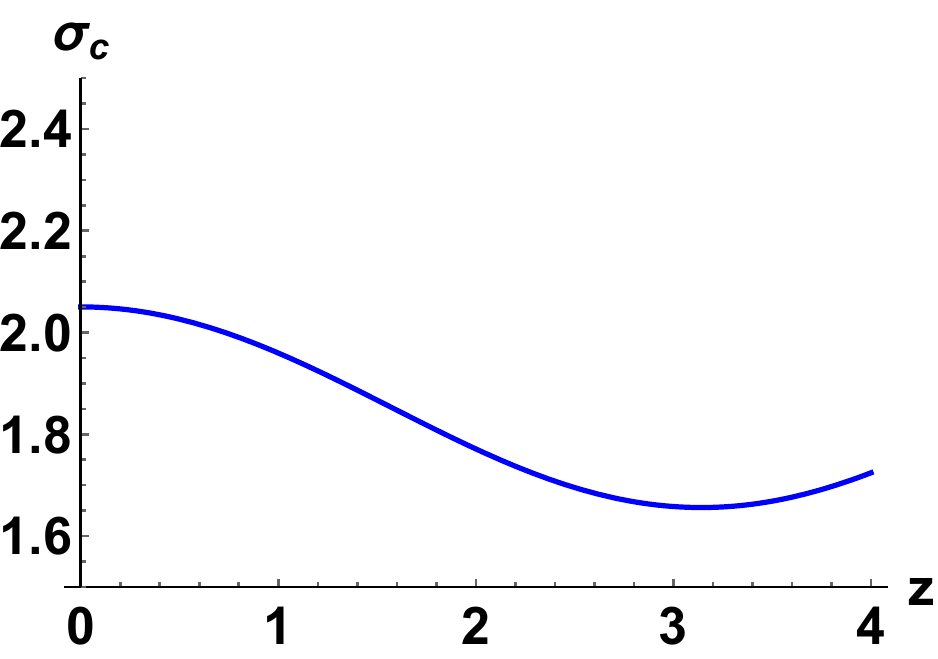} & FeSe$_{1-x}$S$_x$~\cite{dxyex} \\
\hline
\multirow{4}{*}{Topological} & \multirow{2}{*}{\begin{tabular}[c]{@{}c@{}}Chiral\\(Gapful) \end{tabular}} & $ p_x+i p_y $ & Odd & D & $ (k_x+ik_y )\hat{z} $ & $ e^{i \theta} \hat{z} $ & 0 & 0 & \includegraphics[scale=0.1]{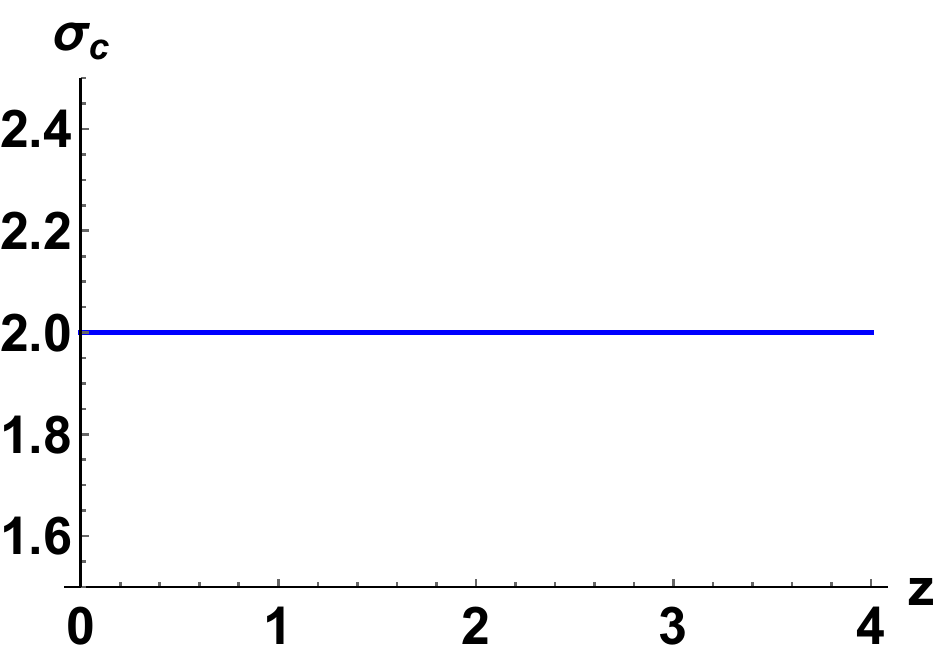} & \begin{tabular}[c]{@{}c@{}}Possibly Sr$_2$RuO$_4$~\cite{Leggett}, \\ UTe$_2$~\cite{chiral-pex}\end{tabular} \\
\cline{3-11}
& & $d_{x^2-y^2}+ i d_{xy} $ & Even & C & 0 & 0 & $k^2_x-k^2_y+i 2 k_x k_y$ & $e^{i 2 \theta} $ & \includegraphics[scale=0.1]{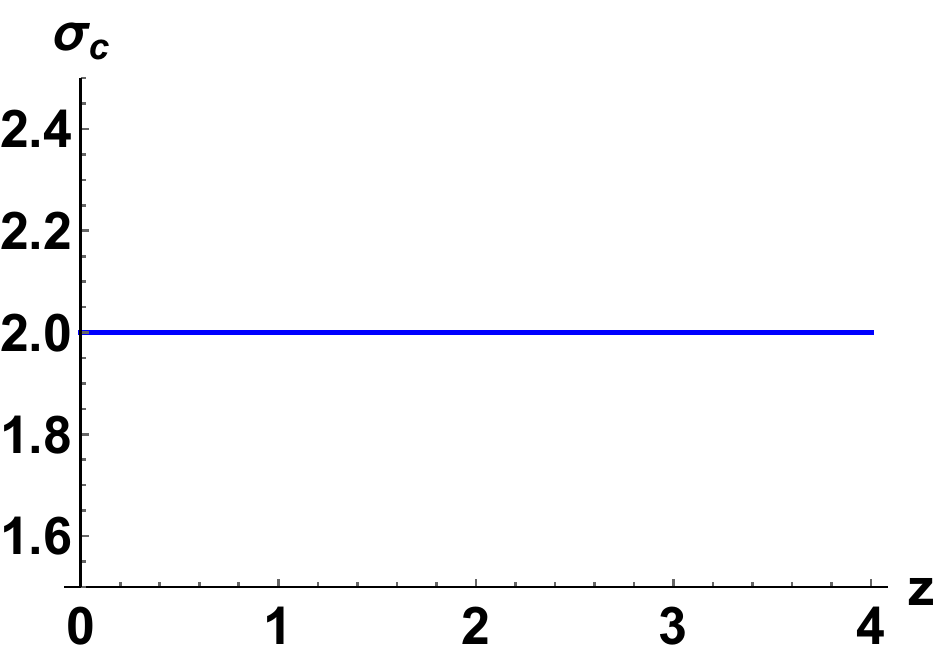} & SrPtAs~\cite{chiraldex} \\
\cline{2-11}
& \begin{tabular}[c]{@{}c@{}}Helical\\(Gapful)\end{tabular} & $p$ & {odd~\cite{parity}} & DIII & $\hat{x} k_y - \hat{y} k_x$ & $\sin{\theta} \hat{x} - \cos{\theta} \hat{y}$ & 0 & 0 & \includegraphics[scale=0.1]{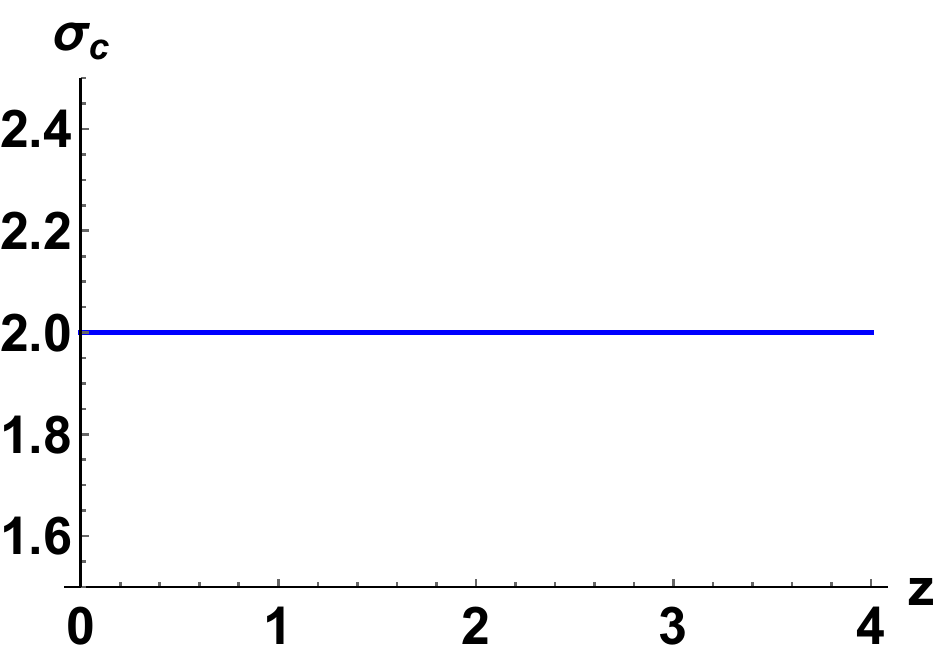} & Possibly Sr$_2$RuO$_4$~\cite{Leggett} \\
\cline{2-11}
& Nodal & $p_x\hat{x}$,~$p_x \hat{y}$ & {odd~\cite{parity}} & DIII & $k_x \hat{x}$,~$k_x \hat{y}$ & $\cos{\theta} \hat{x}$,~$\cos{\theta} \hat{y}$ & 0 & 0 & \includegraphics[scale=0.1]{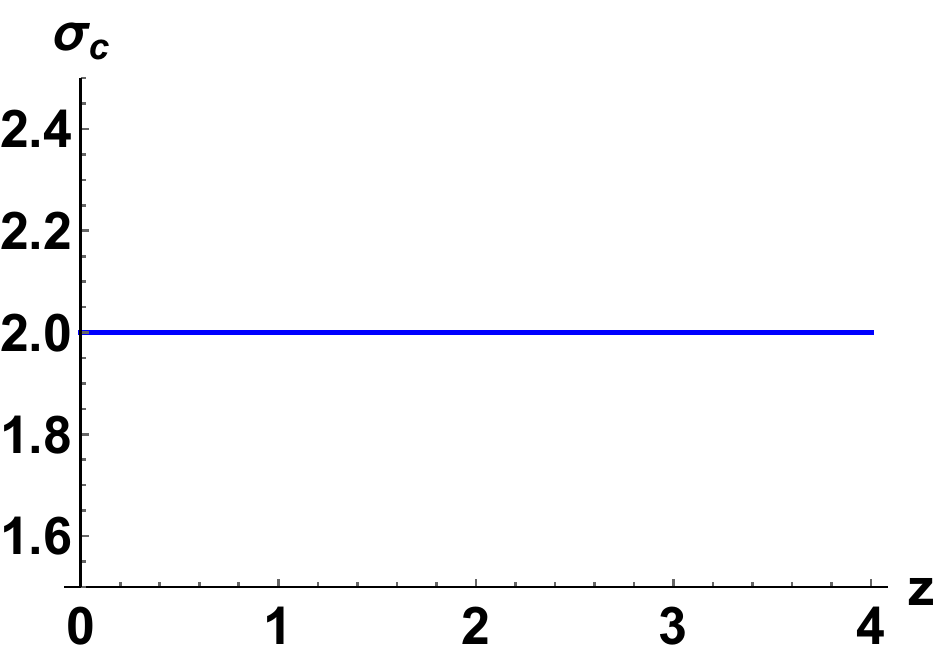} & - \\
\hline
\end{tabular}}
\label{table pairing}
\end{table}

\end{widetext}

The pairing symmetry of superconductors is classified based on spin angular momentum ($S$) of Cooper pairs, which form via pairing between two spin half electrons. It can be either singlet ($0$) or triplet ($1$). Spin-singlet states are antisymmetric in spin space, while spin-triplet states are symmetric. Column 4 in Table ~\ref{table pairing} describes the parity of the different symmetries. {In the absence of spin-orbit coupling, pairing symmetries are either even parity (spin-singlet) or odd parity (spin-triplet). Pairing symmetries, as mentioned in Table~\ref{table pairing}, that satisfy $\psi(\textbf{k}) = \psi(-\textbf{k})$ are associated with spin-singlet (even parity) superconductors, and $\textbf{d}(\textbf{k})=- \textbf{d}(-\textbf{k})$ are associated with spin-triplet (odd parity) superconductors~\cite{parity}. The presence of the spin-orbit coupling term breaks the inversion symmetry. For superconductors with pairing: gapful (helical-$p$) and nodal ($p_x \hat{x}$, $p_x \hat{y}$), there is a spin-orbit coupling term, which too breaks the inversion symmetry.} When the orbital part of the wavefunction is even, it is denoted as even parity with orbital angular momentum ($L=0,2,4...$). When the orbital part is odd, it is called odd parity with orbital angular momentum ($L=1,3,5...$). The total wave function of the Cooper pair must be antisymmetric under the exchange of particles. Thus either the orbital part is antisymmetric, and the spin part is symmetric or vice versa.
Spin-singlet pairing is associated with even orbital angular momentum, and spin-triplet pairing is associated with odd orbital angular momentum~\cite{chiral}. Even parity, spin-singlet Cooper pair states with angular momentum $L= 0,2$ are denoted as $s$-wave, $d$-wave while odd parity, spin-triplet Cooper pair states with angular momentum $L= 1,3$ are denoted as $p$-wave, $f$-wave.

Column 5 in Table~\ref{table pairing} describes ten-fold symmetry classification. Unconventional superconductors, in general, can be categorized as topological or non-topological based on the well-known ten-fold symmetry classification. For 2D superconductors based on Bogoliubov de-Gennes Hamiltonian, pairing symmetries can be categorized into different symmetry classes~\cite{tenfold} based on (1) spin-rotation symmetry, i.e., rotation about $z$ component of spin, by the presence or absence of (2) time-reversal symmetry(TRS), (3) chiral symmetry also known as sublattice symmetry (SLS) and (4) particle-hole symmetry (PHS). For a system where TRS is present, $\hat{T}=\pm 1$ where $\hat{T}$ is the TRS operator, and when TRS is absent, $\hat{T}=0$. Similarly, when PHS is present, $\hat{P}=\pm 1$, $\hat{P}$ being the PHS operator, and when PHS is absent, $\hat{P}=0$. Chiral symmetry can be defined based on whether TRS and PHS are present or not. TRS and PHS together determine chirality or SLS. When SLS is present, $\hat{C}=1$ (where $\hat{C}$ is the chiral symmetry operator), and when absent, $\hat{C}=0$. When either TRS or PHS is absent, SLS is absent $\hat{C}=0$, which includes AI, AII, C, and D symmetry classes. When both TRS and PHS are present, SLS is present $\hat{C}=1$, which includes CI, CII, BDI, and DIII symmetry classes. {For a superconductor, if TRS and PHS are absent, but SLS holds, i.e., $\hat{C}=1$, then it is classified as AIII symmetry, but if SLS is absent, i.e., $\hat{C}=0$, then it is A symmetry class.}\\

We first discuss nodal triplet pairing symmetries, e.g., $p_x \hat{x}$, $p_x \hat{y}$ and $p_x \hat{z}$. Among these, $p_x \hat{x}$ and $p_x \hat{y}$ do not possess spin rotation symmetry but TRS, PHS and SLS are present and therefore $p_x \hat{x}$ and $p_x \hat{y}$ belong to DIII symmetry class which is topological in 2D. {However, when spin rotation symmetry is present in case of a 2D superconductor, there is an exception for AIII symmetry class. AIII symmetry class is then uniquely defined by TRS and spin rotation symmetry in 2D. Those materials for whom both spin rotation symmetry $S_z$ (spin rotation symmetry SU(2) around $z$ direction) and  TRS are preserved, can also be classified as AIII symmetry class. In 2D superconductors, $p_x \hat{z}$ pairing is an example of an AIII symmetry class that possesses spin rotation symmetry and preserves TRS.} This is non-topological in 2D; see Refs.\cite{tenfold, AIII} for a detailed explanation. Nodal singlet pairing, such as $d_{xy}$ and $d_{x^2-y^2}$ preserve TRS and possess spin rotation symmetry. These are examples of 2D BDG Hamiltonian in CI symmetry class that preserves PHS and SLS and thus are non-topological in 2D.

Gapful pairing symmetries such as gapful triplet chiral-$p$ pairing with $\textbf{d}$ parallel to $z$ direction, possess spin rotation symmetry around fixed $z$ axis and preserve PHS but TRS and SLS are absent and thus belong to the D symmetry class in 2D which is topological. Next, gapful singlet chiral$-d$ pairing possesses complete spin rotation symmetry. It preserves PHS, but TRS and SLS are absent and are an example of the C symmetry class in 2D, which is topological. Gapful triplet helical-$p$ pairing preserves TRS and PHS but does not possess spin rotation symmetry, but preserves SLS, and is a member of the DIII symmetry class in 2D, which is categorized as a gapful topological superconductor. Conventional $s$-wave superconductors are of the CI symmetry class. These possess complete spin rotation symmetry and preserve both TRS and PHS, thus preserving SLS, which is non-topological in 2D. {For pairing cases without spin rotation symmetry, there is small but finite Rashba spin-orbit coupling, and for cases with spin rotation symmetry, there is zero Rashba spin-orbit coupling}.

In columns 6-9 of Table~\ref{table pairing}, pairing potential ($\hat{\Delta}(\textbf{k})$) of a Cooper pair is written in terms of $\textbf{d}(\textbf{k})$ for spin-triplet and a scalar term $\psi(\textbf{k})$ for spin-singlet pairing. For spin-triplet pairing such as topological gapful (chiral-$p$, chiral-$d$, helical-$p$) and nodal $p_x$ pairing~\cite{nontrivial, pairing, chiral-p}, pair potential is given as,
\begin{eqnarray}
\hat{\Delta}(\textbf{k})= \Delta (\textbf{d}(\textbf{k}). \hat{\sigma}) i \sigma_2,
\label{eqn:deltap}
\end{eqnarray}

where $\textbf{d}(\textbf{k})$ is defined for each pairing symmetry in Table~\ref{table pairing}, $\hat{\sigma}=\sigma_1 \hat{x} + \sigma_2 \hat{y} + \sigma_3 \hat{z}$ where $\sigma_{1,2,3}$ are the three Pauli matrices and $\Delta$ is the magnitude of the superconducting gap. For 2D superconductor, momentum components are $k_x=k_F \cos \theta$, $k_y=k_F \sin \theta$, where $k_F$ is Fermi wave vector and $\theta$ is angle the incident electron makes with $x$ axis.

$p_x \hat{z}$ pairing is a nodal non-topological superconductors with $\textbf{d}$ in $\hat{z}$ direction, i.e., $\textbf{d}(\textbf{k})=(k_x) \hat{\textbf{z}}/ k_F= \cos \theta \hat{\textbf{z}}$. spin-triplet $p$ wave states or chiral-$p$ pairing, has $\textbf{d}(\textbf{k})=(k_x \pm i k_y) \hat{\textbf{z}}/ k_F$ with $\textbf{d}||\hat{c}$, i.e., $\textbf{d}$ is parallel to crystal $c$ axis~\cite{chiral-pvector}, implying, $\textbf{d}(\textbf{k})=(\cos \theta \pm i \sin \theta) \hat{\textbf{z}}$. In helical-$p$ wave superconductor, $\textbf{d}$ is defined as $(\hat{\textbf{x}} k_x \pm \hat{\textbf{y}} k_y)/ k_F$ or $(\hat{\textbf{x}} k_y \pm \hat{\textbf{y}} k_x)/ k_F$ with $\textbf{d}\perp \hat{c}$ ($\textbf{d}$ in $ab$ plane), i.e., $\textbf{d}$ is perpendicular to the crystal $c$ axis~\cite{helicalp}, implying, $\textbf{d}(\textbf{k})=(\hat{\textbf{x}} \cos \theta \pm \hat{\textbf{y}} \sin \theta)$ or $(\hat{\textbf{x}} \sin \theta \pm \hat{\textbf{y}} \cos \theta)$. For nodal topological $p_x$ pairing, $\textbf{d}$ can be in $\hat{x}$ or $\hat{y}$ direction, i.e., $\textbf{d}(\textbf{k}) =k_x \hat{x}/k_F= \cos (\theta) \hat{x}$ or $\textbf{d}(\textbf{k}) =k_x \hat{y}/k_F= \cos (\theta) \hat{y}$.

For spin-singlet pairing such as non-topological gapful $s$, non-topological nodal $d_{x^2-y^2}$ and $d_{xy}$, topological gapful chiral-$d$ pairing~\cite{pairing, dpairing}, pair potential is given as,
\begin{eqnarray}
\Delta(\textbf{k})= \Delta (\psi(\textbf{k})) ,
\label{eqn:deltad}
\end{eqnarray}
where $\psi(\textbf{k})$ is a scalar term that represents spin-singlet pairing and is defined in Table~\ref{table pairing}. {$s$-wave superconductors are non-topological gapful superconductors~\cite{swave}}. $d_{x^2-y^2}$ pairing is a nodal non-topological superconductor, where the order parameter vanishes, diagonal to $x,y$ directions~\cite{dwave} with $\psi(\textbf{k})= (k^2_x-k^2_y)/k^2_F=\cos 2 \theta$. $d_{xy}$ pairing is a nodal non-topological superconductor, where the order parameter vanishes in $x,y$ direction~\cite{burset} with $\psi(\textbf{k}) =(2 k_x k_y)/k^2_F = 2 \cos \theta \sin \theta$. For chiral $d$ wave superconductor, $\psi(\textbf{k})=((k^2_x-k^2_y)+i 2 k_x k_y)/k^2_F= e^{i 2 \theta}$.

In Table~\ref{table pairing}, column 10, we have plotted normalized conductance ($\sigma_c$) vs. barrier strength ($z$) at zero bias for a metal/2D superconductor junction which is a typical signature of the existence of Majorana bound states at metal/topological superconductor interface. Normalized conductance $\sigma_c$ is quantized for topological superconductors, e.g., chiral-$p$, chiral-$d$, helical-$p$, $p_x \hat{x}$ and $p_x \hat{y}$ cases and is in line with the ten-fold symmetry classification for 2D BDG Hamiltonian~\cite{tenfold}.

In Table~\ref{table pairing} column 11, examples of topological and non-topological superconductors are given with their respective pairing symmetries. Topological gapful superconductors are chiral $p$-wave which are $p_x+$i$p_y$ superconductor~\cite{chiralp} and chiral $d$-wave which is $d_{x^2-y^2}+$i$d_{xy}$ spin-singlet superconductor~\cite{chirald}. In Table ~\ref{table pairing}, we give examples of unconventional superconductors for which the said pairing is most likely. For example, the pairing symmetry of Sr$_2$RuO$_4$ is still inconclusive, as regards chiral-$p$ or helical-$p$~\cite{Leggett}. The next section explains the $2D$ BTK (Blonder, Tinkham, and Klapwijk) approach, and using it, we calculate HBT correlations and non-local conductance for the chosen 2D $N_1$/I/US/I/$N_2$ setup.

\section{2D BTK approach}
\label{2dBTK}

\begin{figure}[h!]
\centering
\includegraphics[scale=0.38]{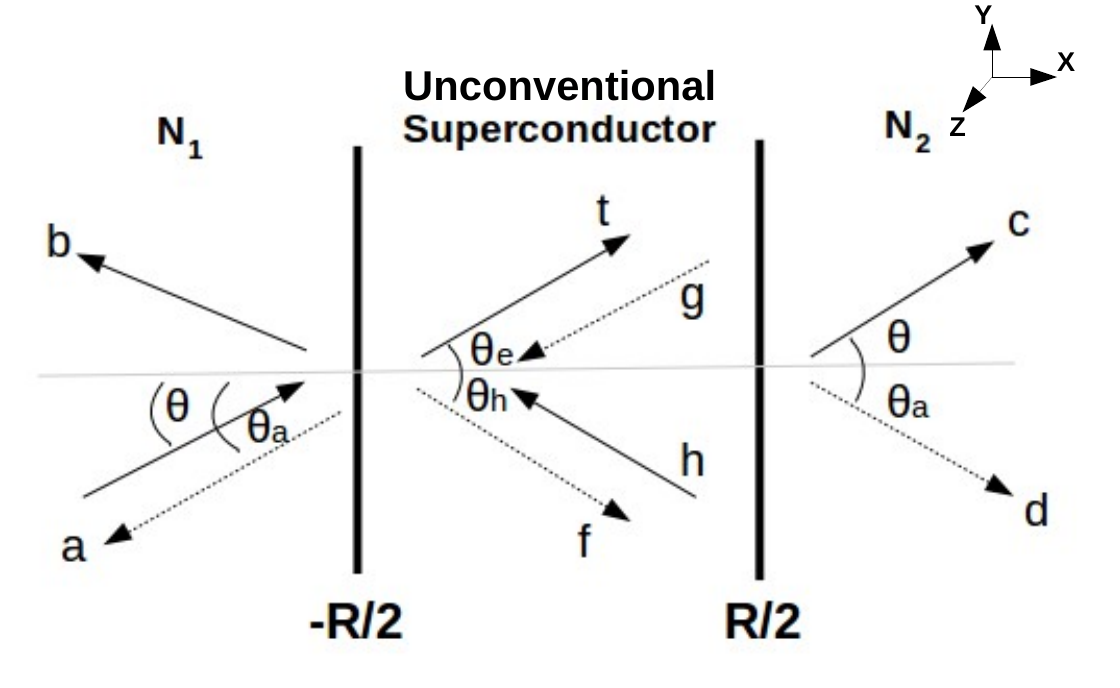}
\caption{Schematic illustration of reflection and transmission of electrons/holes in a normal metal($N_1$)/insulator(I)/unconventional superconductor(US)/insulator(I)/normal metal($N_2$) junction in $x-y$ plane. The solid line represents the scattering of electrons, while the dotted line represents the scattering of holes.}
\label{fig:SCHEMATICNUSN}
\end{figure}
We consider a 2D $N_1$/I/US/I/$N_2$ junction see Fig.~\ref{fig:SCHEMATICNUSN}, with insulators at $x=-R/2$ and $x=R/2$. $\theta$ is angle, the incident (transmitted) electrons (holes) make with $x$ axis~\cite{2DBTK} in $N_1$ ($N_2$) region. $\theta_{a}$ is angle, reflected (transmitted) holes (electrons) make with $x$ axis in $N_1$ ($N_2$) region. Finally, $\theta_{e(h)}$ is angle, transmitted electrons (holes) make with $x$ axis in $US$ region. {We use the Andreev approximation in normal metal regions, which implies that the electron and hole wave vectors are identical and equal to the Fermi wave vector ($k_F$). We consider Andreev approximation in the US region for cases with spin rotational symmetry (in the absence of Rashba spin-orbit coupling), which implies that electron-like quasiparticles and hole-like quasiparticles wave vectors are equal to Fermi wave vector ($k_F$). In contrast, there is no Andreev approximation in the US region for cases without spin rotational symmetry; here, Rashba spin-orbit coupling is finite.} Due to presence of translational invariance in $y$ direction~\cite{2DBTK1}, electron, hole and quasiparticle wavevectors in $y$ direction in $N_1$, $N_2$ and $US$ regions are conserved. This further implies that $\theta=\theta_a=\theta_{e(h)}$.

\begin{figure}[h]
\centering
\includegraphics[scale=0.27]{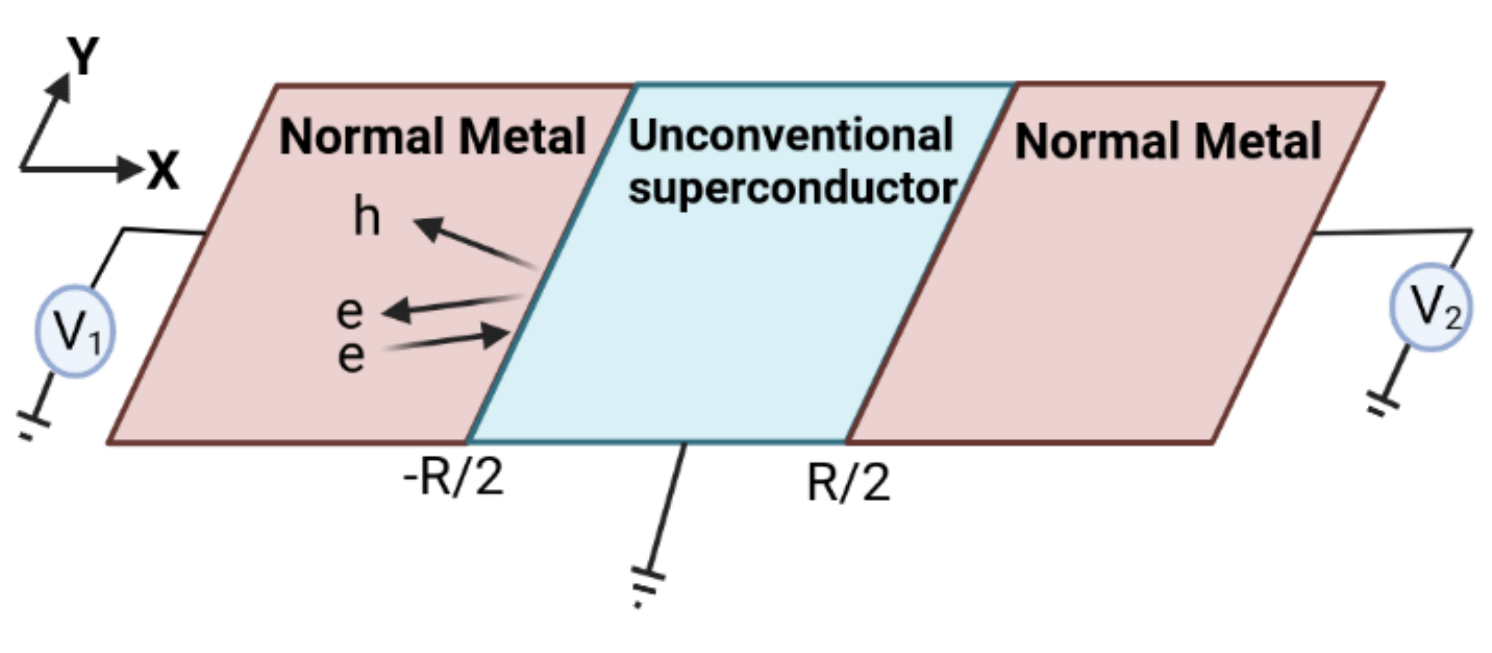}
\caption{2D $N_1$/I/US/I/$N_2$ junction. The superconductor is grounded while bias voltages $V_1$ is applied to $N_1$ and $V_2$ is applied to $N_2$. At $-R/2$, $R/2$ these are insulators which are represented by $\delta$ function potentials.}
\label{fig:NUSN}
\end{figure}

In Fig.~\ref{fig:SCHEMATICNUSN} Andreev reflection amplitude is denoted as $a = s^{eh}_{11}$, normal reflection amplitude as $b = s^{ee}_{11}$, transmission amplitude of elastic cotunneling $c = s^{ee}_{12}$, transmission amplitude of cross Andreev reflection $d=s^{eh}_{12}$. Scattering amplitude $s^{\alpha \gamma}_{i k}$ represents a particle $\alpha$ ($\in e,h$) incident from contact $i$ ($\in N_1,N_2$) which is reflected or transmitted to contact $k$ ($\in N_1,N_2$) as a particle $\gamma$ ($\in e,h$).
In Fig.~\ref{fig:NUSN} we show our chosen setting to study transport and current cross-correlations across unconventional superconductors for a 2D $N_1$/I/US/I/$N_2$ junction wherein $N_1$ is at bias voltage $V_1$ and $N_2$ is at bias voltage $V_2$ while US is grounded.

\subsection{Non-local conductance and shot noise}
Andreev reflection can be non-local, too, known as crossed Andreev reflection(CAR), in which an electron is transmitted as a hole through the other interface with a superconductor, i.e., spatially separated entangled electrons in both the normal metals~\cite{gcar}. When an electron (hole) from the left normal metal tunnels to the right normal metal as an electron (hole), the process is known as elastic co-tunneling (EC).

First, we calculate non-local differential conductance, defined as conductance in $N_2$ when both $N_2$ and $US$ are grounded, and voltage is applied to $N_1$. The difference between CAR and elastic cotunneling (EC) in absence of any voltage bias to $N_2$ is defined as non-local differential conductance~\cite{gcar}, $G_{NL} = G_{CAR}-G_{EC}$, where
\begin{equation}
G_{CAR}= \int^{\pi/2}_{-\pi/2} d\theta \frac{\cos{\theta}}{2 \pi} |d|^2,
~~G_{EC}= \int^{\pi/2}_{-\pi/2} d\theta \frac{\cos{\theta}}{2 \pi} |c|^2,
\label{eqn:G}
\end{equation}{}
$d=s^{eh}_{12}$ is scattering amplitude for CAR while $c=s^{ee}_{12}$ is scattering amplitude for EC.
Shot noise can give more information regarding the pairing symmetry from current cross-correlations in the $N_1$/I/US/I/$N_2$ junction, which conductance cannot. It is why we look at cross-correlations and differential shot noise cross-correlations. Discreteness of charge leads to non-equilibrium temporal fluctuations in the current known as shot noise. Even at zero temperature, shot noise exists, unlike thermal noise that originates due to finite temperature and vanishes at zero temperature~\cite{buttiker}.

The general result for shot noise cross-correlations using 2D BTK for our setup, see Fig.~\ref{fig:NUSN}, includes an integral over incident angle $\theta$ and is given~\cite{datta, noise} as,
\begin{eqnarray}
S^{12} && = \frac{2e^2}{h} \int^{\pi/2}_{-\pi/2} d\theta \frac{\cos{\theta}}{2 \pi} \sum_{ \substack{k,l \in 1, 2 ,\\
x,y,\gamma,\delta \in e,h} } sgn(x) sgn(y) \int W_{k,\gamma;l,\delta}(1x,E) \nonumber \\
&& W_{l,\delta;k,\gamma}(2y,E) \textit{f}_{k \gamma}(E) [1-\textit{f}_{l \delta}(E)] dE,
\label{eqn:sn}
\end{eqnarray}{}
where parameter $W_{k,\gamma;l,\delta}(1x,E) = \delta_{1 k} \delta_{1 l} \delta_{x \gamma} \delta_{x \delta} - s^{x \gamma *}_{1k}(E) s^{x \delta}_{1 l}(E)$ contains information about the scattering process. $s^{x \gamma}_{1 k}(E)$ represents scattering amplitude, with $\gamma$ denoting an electron or hole incident from contact $k$ which is transmitted to $N_1$ as particle of type $x$. $\textit{f}_{k \gamma}$ is Fermi function for particle of type $\gamma$ in contact $k$. Normal metal $N_1$ is contact $1$ while normal metal $N_2$ is contact $2$. Here $sgn(x)=+1$ for $x=e$, i.e, electron and $sgn(x)=-1$ for $x=h$, i.e., hole.
In a previous study~\cite{floser} it was shown that sign of cross-correlations and sign of differential shot noise cross-correlations could be different for some regime of bias voltages applied. Taking a cue from this, we study both shot noise cross-correlations and differential shot noise cross-correlations. It will help us understand different pairing symmetries in our setup.
The differential shot noise cross-correlations $\frac{dS^{12}}{dV}$ in symmetric setup ($V_1=V_2=V$) at zero temperature is given as~\cite{datta, floser},
\begin{equation}
\frac{dS^{12}(V_1=V_2=V)}{dV} = \frac{4|e|^3}{h} sgn(|e|V) \int^{\pi/2}_{-\pi/2} d\theta \frac{\cos{\theta}}{2 \pi} \left(- s_A + s_B \right),
\label{eqn:dsn sym}
\end{equation}
where $s_A= \sum_{\alpha=a,b,c,d}(s_\alpha (|e|V)+s_\alpha (-|e|V))$, with,
\begin{align}
& s_a = s^{eh}_{21}s^{he}_{12}s^{hh *}_{11}s^{ee *}_{22} + s^{eh}_{12}s^{he}_{21}s^{ee *}_{11 }s^{hh *}_{22}, \nonumber \\
&s_b = s^{hh}_{12}s^{ee}_{21}s^{eh *}_{22}s^{he *}_{11} + s^{ee}_{12}s^{hh}_{21}s^{eh *}_{11}s^{he *}_{22}, \nonumber \\
& s_c = s^{hh}_{11}s^{ee}_{21}s^{eh *}_{21}s^{he *}_{11} + s^{ee}_{11} s^{hh}_{21}s^{he *}_{21}s^{eh *}_{11}, \nonumber \\
&s_d = s^{eh}_{12} s^{he}_{22} s^{ee *}_{12} s^{hh *}_{22} + s^{hh}_{12} s^{ee}_{22} s^{he *}_{12} s^{eh *}_{22}, \nonumber
\end{align}
and $s_B= \sum_{\alpha=e,f,g,h} (s_\alpha (|e|V)+s_\alpha (-|e|V))$, with\\
\begin{align}
& s_e = s^{ee}_{12} s^{eh}_{21} s^{ee *}_{22} s^{eh *}_{11} + s^{hh}_{21} s^{he}_{12} s^{hh *}_{11} s^{he *}_{22}, \nonumber \\
&s_f = s^{eh}_{12} s^{ee}_{21} s^{ee *}_{11} s^{eh *}_{22} + s^{hh}_{12} s^{he}_{21} s^{hh *}_{22} s^{he *}_{11},\nonumber \\
& s_g = s^{eh}_{11} s^{ee}_{21} s^{ee *}_{11} s^{eh *}_{21} + s^{hh}_{21} s^{he}_{11} s^{hh *}_{11} s^{he *}_{21}, \nonumber \\
&s_h = s^{ee}_{12} s^{eh}_{12} s^{ee *}_{22} s^{eh *}_{12} + s^{eh}_{12} s^{he}_{22} s^{hh *}_{22} s^{he *}_{12}.\nonumber
\end{align}
The differential shot noise cross-correlations in non-local setup at zero temperature with bias voltage $V_1=V$ applied to N$_1$ while N$_2$ grounded is given by,
\begin{equation}
\frac{dS^{12}(V_1=V,V_2=0)}{dV} = \frac{4 |e|^3}{h} sgn(|e|V) \int^{\pi/2}_{-\pi/2} d\theta \frac{\cos{\theta}}{2 \pi} \left(- s_C + s_D \right),
\label{eqn:dsn asym}
\end{equation}
where $s_C= s_m(-|e|V)+s_n(|e|V)+s_c(|e|V)+s_c(-|e|V)+s_i(|e|V)+s_j(-|e|V)$, and $s_D=s_e(|e|V)+ s_g(|e|V)+s_g(-|e|V)+s_k(|e|V)+s_l(-|e|V)+s_f(-|e|V)$ with,
\begin{align}
&s_m = s^{eh}_{21} s^{he}_{12} s^{hh *}_{11} s^{ee *}_{22} + s^{ee}_{12} s^{hh}_{21} s^{eh *}_{11} s^{he *}_{22}, \nonumber \\
&s_n = s^{eh}_{12} s^{he}_{21} s^{ee *}_{11 } s^{hh *}_{22} + s^{hh}_{12} s^{ee}_{21} s^{eh *}_{22} s^{he *}_{11}, \nonumber \\
&s_i = s^{ee}_{12} s^{he}_{21} s^{ee *}_{11} s^{he *}_{22} + s^{ee}_{21} s^{he}_{12} s^{ee *}_{22} s^{he *}_{11}, \nonumber \\
& s_j = s^{eh}_{12} s^{hh}_{21} s^{hh *}_{22} s^{eh *}_{11} + s^{hh}_{12} s^{eh}_{21} s^{hh *}_{11} s^{eh *}_{22}, \nonumber \\
&s_k = s^{ee}_{12} s^{ee}_{21} s^{ee *}_{11} s^{ee *}_{22} + s^{he}_{21} s^{he}_{12} s^{he *}_{11} s^{he *}_{22}, \nonumber \\
& s_l = s^{hh}_{12} s^{hh}_{21} s^{hh *}_{11} s^{hh *}_{22} + s^{eh}_{12} s^{eh}_{21} s^{eh *}_{11} s^{eh *}_{22}.\nonumber
\end{align}
The non-local conductance and shot noise results for normal incidence ($\theta=0$) agree with 1D BTK results~\cite{BTK}. In the next section, we write the wave functions and boundary conditions for the $N_1$/I/US/I/$N_2$ junctions, first for pairing symmetries that do not possess spin rotation symmetries, followed by pairing symmetries that possess spin rotation symmetry.

\section{Theory}
\label{theory}

We first discuss the wavefunctions for pairing symmetries with spin rotation symmetry, i.e., $s$-wave, $d_{xy}$, $d_{x^2-y^2}$, $p_x \hat{z}$, chiral-$p$ and chiral-$d$ followed by their boundary conditions. Next, we discuss wavefunctions for pairing symmetries that do not possess spin rotation symmetry, i.e., $p_x \hat{x}$, $p_x \hat{y}$ and helical-$p$ followed by their boundary conditions.\\

\subsection{With spin rotation symmetry}

{The 2D Hamiltonian for BDG equation $\mathcal{H} \psi= E \psi$, for US which possesses full spin-rotation symmetry, i.e., for pairing symmetries $d_{xy}$, $d_{x^2-y^2}$, $s$-wave and chiral-$d$ pairing and finally with fixed spin rotation symmetry around $z$ axis, i.e., $p_{x} \hat{z}$, chiral-$p$ pairing, is a $2 \times 2$ matrix~\cite{tenfold,chiral-p}, and is written as,}
\begin{equation}
\mathcal{H}=\left(
\begin{array}{cc}
H_0(\textbf{k}) & \Delta(\textbf{k}) \\
\Delta^{\dagger}(\textbf{k}) & -H^*_0(-\textbf{k})
\end{array}
\right),
\label{eqn:Hamwspin}
\end{equation}

{with eigen spinor ($c^{\dagger}_{k},c_{-k})$, where $c^{\dagger}_{k}(c_{-k})$ being the creation (annihilation) operators.} Pairing potential for triplet pairing is $\Delta(\textbf{k})= \Delta d_z(k)$ for $p_x \hat{z}$ and chiral-$p$ cases as defined in Eq.~\ref{eqn:deltap}, and in terms of $\psi(\textbf{k})$ for singlet as $\Delta(\textbf{k})= \Delta \psi(k)$, i.e., $d_{xy}$, $d_{x^2-y^2}$, $s$-wave, chiral-$d$ cases is defined in Eq.~(\ref{eqn:deltad}). {Pairing symmetries that satisfy $\Delta(\textbf{k}) = \Delta(-\textbf{k})$ are spin-singlet (even parity) superconductors~\cite{parity} for cases with spin rotation symmetry in the absence of Rashba spin-orbit coupling.} In Eq.~(\ref{eqn:Hamwspin}), $H_0 (\textbf{k})=\left( -\dfrac{\hbar^2 \textbf{k}^2} { 2 m} +U(x)-E_F \right)$, wave functions in $N_1$, US and $N_2$ regions for $d_{xy}$, $d_{x^2-y^2}$, $s$-wave, chiral-$p$, chiral-$d$ and $p_{x} \hat{z}$ pairing, and for an electron incident from $N_1$ are given as,
\begin{widetext}
\begin{align}
\psi_{N_1}(x) &=e^{i k_F y \sin{\theta}} \left[ \left( \begin{array}{c}
1\\
0\\
\end{array} \right) (e^{i k_x x } + b e^{-i k_x x } ) +
a \left( \begin{array}{c}
0\\
1\\
\end{array} \right) e^{i k_x x} \right] ,~\text{for}~x < -\frac{R}{2}, \nonumber \\
\psi_{US}(x) &= e^{i k_F y \sin{\theta}} \Biggl[ \left( \begin{array}{c}
u(\theta) \\
\eta^*(\theta) v(\theta) \\
\end{array} \right) t e^{i k_x (x+\frac{R}{2})} e^{-(x+\frac{R}{2})/ \xi}+
\left( \begin{array}{c}
\eta(\theta) v(\theta) \\
u(\theta)\\
\end{array} \right) f e^{-i k_x (x+\frac{R}{2})} e^{-(x+\frac{R}{2})/ \xi} + \left( \begin{array}{c}
u(\theta_-) \\
\eta^*(\theta_-) v(\theta_-) \\
\end{array} \right) \nonumber \\
& g e^{-i k_{x} (x-\frac{R}{2})} e^{(x-\frac{R}{2})/ \xi} +
\left( \begin{array}{c}
\eta(\theta_-) v(\theta_-) \\
u(\theta_-) \\
\end{array} \right) h e^{i k_{x} (x-\frac{R}{2})} e^{(x-\frac{R}{2})/ \xi} \Biggr], ~\text{for}~ -\frac{R}{2} < x < \frac{R}{2}, ~\text{and} \nonumber \\ \psi_{N_2}(x) &= e^{i k_F y \sin{\theta}} \left[ c \left( \begin{array}{c}
1\\
0
\end{array} \right) e^{i k_{x} (x-\frac{R}{2})} +
d \left( \begin{array}{c}
0\\
1
\end{array} \right) e^{-i k_{x} (x-\frac{R}{2})} \right],~ \text{for} \hspace{0.09cm} x > R/2,
\label{eqn:wf wspin}
\end{align}
\end{widetext}

where $\eta(\theta_{\pm})= \Delta(\theta_{\pm})/|\Delta(\theta_{\pm})|$ with $\theta_+=\theta$ and $\theta_-=\pi - \theta$. $\Delta(\textbf{k},x)=\Delta$ from Eq.~\ref{eqn:deltad} is constant in $s$-wave superconducting region and is zero in the normal metal regions. For gapful chiral-$p$ superconductor~\cite{chiral-p}, pair potential is $\Delta(\theta_\pm)=\Delta e^{i \theta_{\pm}}$ and for gapful chiral-$d$ superconductor, $\Delta(\theta_\pm)=\Delta e^{2 i \theta_{\pm}}$. For nodal $d_{x^2-y^2}$ superconductor, pairing potential is $\Delta(\theta_\pm)=\Delta \cos(2 \theta_{\pm})= \Delta \cos(2 \theta)$, for $d_{xy}$ superconductor, $\Delta(\theta_\pm)=2 \Delta \cos(\theta_{\pm}) \sin(\theta_{\pm})=\pm2 \Delta \cos(\theta) \sin(\theta)$, and for $p_{x} \hat{z}$ superconductor, $\Delta(\theta_\pm)= \Delta \cos(\theta_{\pm})=\pm \Delta \cos(\theta)$. The coherence factors are $u(\theta_{\pm})=\sqrt[]{(E+ \sqrt[]{E^2-|\Delta(\theta_{\pm})|^2})/(2 E)}$ and $v(\theta_{\pm})=\sqrt[]{(E - \sqrt[]{E^2-|\Delta(\theta_{\pm})|^2})/(2 E)}$.

\subsubsection{Boundary conditions}
For pairing symmetries that posses spin rotation symmetry, i.e., $\textbf{d}$ is in $z$ direction with finite $d_z$ for spin-triplet superconductors, e.g., $p_x \hat{z}$, chiral-$p$ and for full spin rotation symmetry with scalar $\psi$ for spin-singlet superconductors, e.g., $s$-wave, $d_{xy}$, $d_{x^2-y^2}$ and chiral-$d$, pairing potential is a scalar term. The continuity equation and current conservation  at  interface  lead to boundary conditions as mentioned in Eq.~\ref{eqn:BC with srs}, solving these one can calculate the scattering amplitudes~\cite{boundarycond}.
The general boundary conditions at the interfaces for a $N_1$/I/US/I/$N_2$ junction, where US satisfies spin rotation symmetry at $x=-R/2$ and $R/2$ are given by,

\begin{eqnarray}
&& \Psi_{N_1}|_{x=-R/2} = \Psi_{US}|_{x=-R/2},~~ \Psi_{US}|_{x=R/2} = \Psi_{N_2}|_{x=R/2}, \nonumber \\
&& \frac{\partial}{\partial x}(\Psi_{US} - \Psi_{N_1})|_{x=-R/2}= \left( 2m U_1/\hbar^2 \right) \Psi_{N_1}|_{x=-R/2},\nonumber\\
&& \frac{\partial}{\partial x}(\Psi_{N_2} - \Psi_{US})|_{x=R/2}= \left( 2m U_2/\hbar^2 \right) \Psi_{US}|_{x=R/2}.
\label{eqn:BC with srs}
\end{eqnarray}

The barrier strength at the interface in both Eq.~(\ref{eqn:BC without srs}) and (\ref{eqn:BC with srs}) are characterized by dimensionless parameters $z_i= 2m U_i/\hbar^2 k_F, i = 1,2$. From the scattering amplitudes $a=s^{eh}_{11}, b=s^{ee}_{11}, c=s^{ee}_{12}, d=s^{eh}_{12}$, we obtain Andreev and normal reflection probabilities as $A=|a|^2$ and $B=|b|^2$. $C=|c|^2$ and $D=|d|^2$ define probabilities for electron co-tunneling (EC) and crossed Andreev reflection (CAR) respectively. This paper considers interface barrier strengths $z=z_{1}=z_{2}$.

\subsection{Without spin rotation symmetry}

We adopt Bogoliubov-de Gennes (BDG) approach to study the transport in $N_1$/I/US/I/$N_2$ junction. {The 2D Hamiltonian for BDG equation $\mathcal{H} \psi= E \psi$ without spin-rotation symmetry in case of $p_{x} \hat{x}$, $p_{x} \hat{y}$ and helical-$p$ pairing~\cite{tenfold} can be written as, }
\begin{equation}
\mathcal{H}=\left(
\begin{array}{cc}
H(\textbf{k}) & \hat{\Delta}(\textbf{k}) \\
\hat{\Delta}^{\dagger}(\textbf{k}) & -H^*(-\textbf{k})
\end{array}
\right) ,
\label{eqn:Hamwospin}
\end{equation}

{with eigen spinor $(c^{\dagger}_{k \uparrow},c^{\dagger}_{k \downarrow},c_{-k \uparrow},c_{-k \downarrow})$ of $\mathcal{H}$, where $c^{\dagger}_{k \uparrow}(c_{-k \uparrow})$ denotes creation (annihilation) operator of spin up quasiparticle and $c^{\dagger}_{k \downarrow}(c_{-k \downarrow})$ denotes creation (annihilation) operator of spin down quasiparticle.} Pairing potential $\hat{\Delta}(k)$ for spin-triplet cases ($p_x \hat{x}$, $p_x \hat{y}$ and helical-$p$) is defined via Eq.~(\ref{eqn:deltap}) and $ H(\textbf{k})= H_0(\textbf{k}).\sigma_0 + H_p(\textbf{k})$ with $H_0 (\textbf{k})=\left( -\dfrac{\hbar^2 \textbf{k}^2} { 2 m} +U(x)-E_F \right)$, $H_p(\textbf{k})= \textbf{V(k)}.\hat{\sigma}$ denotes the spin-orbit coupling term and $\textbf{k}=(k_{x },k_{y},0)$. Here $\sigma_0$ is $2 \times 2$ identity matrix and $\hat{\sigma}=\sigma_1 \hat{x} + \sigma_2 \hat{y} + \sigma_3 \hat{z}$ where $\sigma_{1,2,3}$ are the three Pauli matrices. As in ~\cite{helicalp}, $\textbf{V(k)} = \lambda (\hat{x} k_y - \hat{y} k_x)$ with Rashba spin-orbit coupling constant $\lambda$. In $H_0$, $U(x)=U_1 \delta(x+R/2) + U_2 \delta(x-R/2)$ with $U_1$ and $U_2$ being the barrier strengths and $R$ is the thickness in $x$ direction of 2D US lying in $x-y$ plane. The excitation energy $E$ is measured relative to Fermi energy $E_F$, and $m$ is the electron-like or hole-like quasiparticle mass. For simplicity, we neglect self-consistency of the spatial distribution of the pair potential in the US. {In the presence of Rashba spin-orbit coupling for $\lambda \neq 0$, the additional term $H_p(\textbf{k})$ in Hamiltonian breaks the inversion symmetry, i.e., $H_p(\textbf{k})= -H_p (-\textbf{k})$ and $\hat{\Delta}(\textbf{k})=-\hat{\Delta}(-\textbf{k})$ for odd parity superconductors, i.e., helical-$p$, $p_x \hat{x}$ and $p_x \hat{y}$. However, mixed parity states like Noncentrosymmetry superconductors (NCS), e.g., helical-$p$+$s$ pairing, have no definite parity~\cite{NCS}, i.e., $\hat{\Delta}(\textbf{k}) \neq \pm \hat{\Delta}(-\textbf{k})$.}

The wave functions in $N_1$, US and $N_2$ regions, for an electron incident from $N_1$ are
\begin{widetext}
\begin{align}
\psi_{N_1}(x) &=e^{i k_y y} \left[ \left( \begin{array}{c}
1\\
0\\
0\\
0\\
\end{array} \right) (e^{i k_x x} + b_{1} e^{-i k_x x} ) +
b_{2} \left( \begin{array}{c}
0\\
1\\
0\\
0\\
\end{array} \right) e^{- i k_x x}+
a_{1} \left( \begin{array}{c}
0\\
0\\
1\\
0\\
\end{array} \right) e^{i k_x x}+
a_{2} \left( \begin{array}{c}
0\\
0\\
0\\
1\\
\end{array} \right) e^{i k_x x} \right] ,~ \text{for} ~x < -\frac{R}{2}, \nonumber \\
\psi_{US}(x) &= e^{i k_y y} \left[ \left( t_{1} \left( \begin{array}{c}
u \\
-i \alpha^{-1}_1 u \\
i \alpha^{-1}_1 v \\
v \\
\end{array} \right) e^{i k^e_{\uparrow x} (x+\frac{R}{2})} + t_{2} \left( \begin{array}{c}
u \\
i \alpha^{-1}_2 u \\
i \alpha^{-1}_2 v \\
-v \\
\end{array} \right) e^{i k^e_{\downarrow x} (x+\frac{R}{2})} + f_{1} \left( \begin{array}{c}
i \alpha_1 v \\
v \\
u \\
- i \alpha_1 u \\
\end{array} \right) e^{-i k^h_{\uparrow x} (x+\frac{R}{2})} + f_{2} \left( \begin{array}{c}
i \alpha_2 v \\
-v \\
u \\
i \alpha_2 u \\
\end{array} \right) e^{-i k^h_{\downarrow x} (x+\frac{R}{2})} \right) e^{-(x+\frac{R}{2})/ \xi} \right. \nonumber \\
& + \left. \left( g_{1} \left( \begin{array}{c}
u \\
-i \alpha^{-1}_1 u \\
i \alpha^{-1}_1 v \\
v \\
\end{array} \right) e^{-i k^e_{\uparrow x} (x+\frac{R}{2})} + g_{2} \left( \begin{array}{c}
u \\
i \alpha^{-1}_2 u \\
i \alpha^{-1}_2 v \\
-v \\
\end{array} \right) e^{-i k^e_{\downarrow x} (x+\frac{R}{2})} +
h_{1} \left( \begin{array}{c}
i \alpha_1 v \\
v \\
u \\
- i \alpha_1 u \\
\end{array} \right) e^{i k^h_{\uparrow x} (x+\frac{R}{2})} + h_{2} \left( \begin{array}{c}
i \alpha_2 v \\
-v \\
u \\
i \alpha_2 u \\
\end{array} \right) e^{i k^h_{\downarrow x} (x+\frac{R}{2})} \right) e^{(x-\frac{R}{2})/ \xi} \right], ~\text{for} ~-\frac{R}{2} < x < \frac{R}{2}, \nonumber \\
\psi_{N_2}(x) &= e^{i k_y y} \left[ c_{1} \left( \begin{array}{c}
1\\
0\\
0\\
0\\
\end{array} \right) e^{i k_x (x-\frac{R}{2})} +
c_{2} \left( \begin{array}{c}
0\\
1\\
0\\
0\\
\end{array} \right) e^{i k_x (x-\frac{R}{2})}+
d_{1} \left( \begin{array}{c}
0\\
0\\
1\\
0\\
\end{array} \right) e^{-i k_x (x-\frac{R}{2})}+
d_{2} \left( \begin{array}{c}
0\\
0\\
0\\
1
\end{array} \right) e^{-i k_x (x-\frac{R}{2})} \right],~\text{ for} \hspace{0.09cm} x > \frac{R}{2},
\label{eqn:wf wospin}
\end{align}
\end{widetext}

with coherence factors for electron (hole) quasiparticles are given by $u(v)=\sqrt[]{ (E+(-) \sqrt[]{E-| \Delta|^2})/2 E}$. {$\alpha_{1(2)}=k^{h}_{\uparrow(\downarrow)}/k_{\uparrow(\downarrow)}$ where $k_{\uparrow(\downarrow)}=+(-)(m \lambda / \hbar^2)+ \sqrt{(m \lambda / \hbar^2)^2+k^2_F}$, $k^e_{\uparrow(\downarrow)} = k_{\uparrow(\downarrow)} e^{ i \theta_{\uparrow(\downarrow)}}$ and $k^h_{\uparrow(\downarrow)} = k_{\uparrow(\downarrow)} e^{- i \theta_{\uparrow(\downarrow)}}$. $\theta_{\uparrow(\downarrow)}$ denotes phase of the wave with wave number $k_{\uparrow(\downarrow)}$, see Ref.~\cite{helicalp}. As translational symmetry is preserved for $y$ direction, $k_y=k_F \sin \theta=k_{\uparrow} \sin \theta_{\uparrow}=k_{\downarrow} \sin \theta_{\downarrow}$. $k^{e(h)}_{\uparrow(\downarrow)x}$ denotes the $x$ component of wave vector $k^{e(h)}_{\uparrow(\downarrow)x}$ and is defined as $k^{e(h)}_{\uparrow(\downarrow)x} = \sqrt{(k^{e(h)}_{\uparrow(\downarrow)})^2-(k_y)^2}$.}
$a_{1(2)}$ and $b_{1(2)}$ represent scattering amplitudes for Andreev reflection and normal reflection of spin-up(down) quasiparticles. $c_{1(2)}$ and $d_{1(2)}$ represent scattering amplitudes for elastic cotunneling and cross Andreev reflection of spin-up(down) quasiparticles. Superconducting coherence length $\xi=\hbar v_F / \Delta$ where $v_F$ is the Fermi velocity~\cite{xi}.

\subsubsection{Boundary conditions}
For pairing symmetries, $p_{x} \hat{x}$, $p_{x} \hat{y}$ and helical-$p$, that do not possess spin rotation symmetry, and are not spin-polarized, i.e., $\textbf{d}$ is not in one fixed spin direction but in both $x$ and $y$ directions, this results in finite diagonal terms in pairing potential matrix $\hat{\Delta}$. {The scattering amplitudes are determined via the continuity equation and current conservation, which leads to boundary conditions as in Eq.~(\ref{eqn:BC without srs}), see also Ref.~\cite{boundarycond}.} The general boundary conditions at the interfaces for US without spin rotation symmetry, for a $N_1$/I/US/I/$N_2$ junction at $x=-R/2$ and $R/2$ are given by:
\begin{align}
\Psi_{N_1}|_{x=-R/2} = \Psi_{US}|_{x=-R/2},~~ \Psi_{US}|_{x=R/2} = \Psi_{N_2}|_{x=R/2}, \nonumber \\
\hbar v_{USx} \Psi_{US}|_{x=-R/2} - \hbar v_{Nx} \Psi_{N_1}|_{x=-R/2}= -2i U(x) \tau_3 \psi_{US}|_{x=-R/2}, \nonumber \\
\hbar v_{Nx} \Psi_{N_2}|_{x=R/2} - \hbar v_{USx} \Psi_{US}|_{x=R/2}= -2i U(x) \tau_3 \psi_{US}|_{x=R/2},
\label{eqn:BC without srs}
\end{align}
where velocity operator in $x$ direction~\cite{helicalp} is defined by $\hbar v_{USx} = \partial H/ \partial k_x$ for US and $\hbar v_{Nx} = \partial H_{N}/ \partial k_x$ where the Hamiltonian $H_{N}$ for both normal metals $N_1$ and $N_2$ is given by putting the pairing potential $\hat{\Delta}(\textbf{k})=0$ in Eq.~(\ref{eqn:Hamwospin}) and diagonal matrix $\tau_3$ is given by $diag(1,1,-1,-1)$.

\section{Results and Discussion}
Herein, we calculate differential non-local conductance $G_{NL}$, differential shot noise cross-correlations as well as shot noise cross-correlations for the 2D $N_1$/I/US/I/$N_2$ set up as shown in Fig.~\ref{fig:NUSN}, see~\cite{codes} for detailed calculations. We take two cases for bias voltages applied in $N_1$ and $N_2$, i.e., $V_1=$ $V_2$ (symmetric set up) and $V_1 \neq 0$, $V_2=0$ (non-local set up). We plot non-local conductance~\cite{Rbyxivalue, NSN} and differential shot noise with the propagating phase $k_F R=55$ and length of the superconductor in terms of superconducting coherence length $R/\xi=2$. We take these values for propagating phase $k_F R$ and the superconductor's length because when $R >> \xi$, there will be no possibility of non-local transport as electron-like or hole-like quasiparticles. These cannot transmit to the normal metal $N_2$, and when $R << \xi$, the effect of US will be suppressed. However, when the length of superconductor $R$ is comparable to superconducting coherence length $\xi$, incident quasiparticles can transmit to normal metal $N_2$ as an electron(EC) or hole(CAR). From Refs. ~\cite{been,Dong}, one sees a wide range for the propagating phase ($k_F R=10-5000$). It has also been shown that non-local transport will be suppressed for small values~\cite{kfr} of $k_F R$. In spin-triplet topological superconductor (Sr$_2$RuO$_4$)~\cite{chiral-xi}, superconducting coherence length $\xi = 91nm$ and superconducting gap $\Delta= 1.76 ~k_B T_c$, where $k_B$ is Boltzmann constant and critical temperature $T_c=1.6~$K. Thus, we get a value for Fermi wave vector $k_F=3 \times 10^8 m^{-1}$. In this paper, we take the length of US, $R=2 \xi$. This gives us a value for the propagating phase, i.e., $k_F R \simeq 55$. Hence in this paper, we take $k_F R$ to be $55$. {For pairing symmetries without spin rotational symmetry, i.e. $p_x \hat{x}$, $p_x \hat{y}$ and helical-$p$, we have considered $2 m \lambda / \hbar^2 =0.1 k_F$ as has also been taken in Ref.~\cite{helicalp}. Substituting the value of $k_F$, $m$ the mass of electron and $\hbar$, we get $\lambda= 0.17 \times 10^{-30} m$. This $\lambda$ value corresponds to $2 m \lambda / \hbar^2 =0.1 k_F$, implying normalised Rashba spin-orbit strength is around $0.1$ times the Fermi wave vector $k_F$.} The next subsection provides results for differential non-local conductance and shot noise cross-correlations for different pairing symmetries.

\subsection{Differential non-local conductance and differential shot noise cross-correlations}
\begin{figure}[h!]
\includegraphics[scale=0.32]{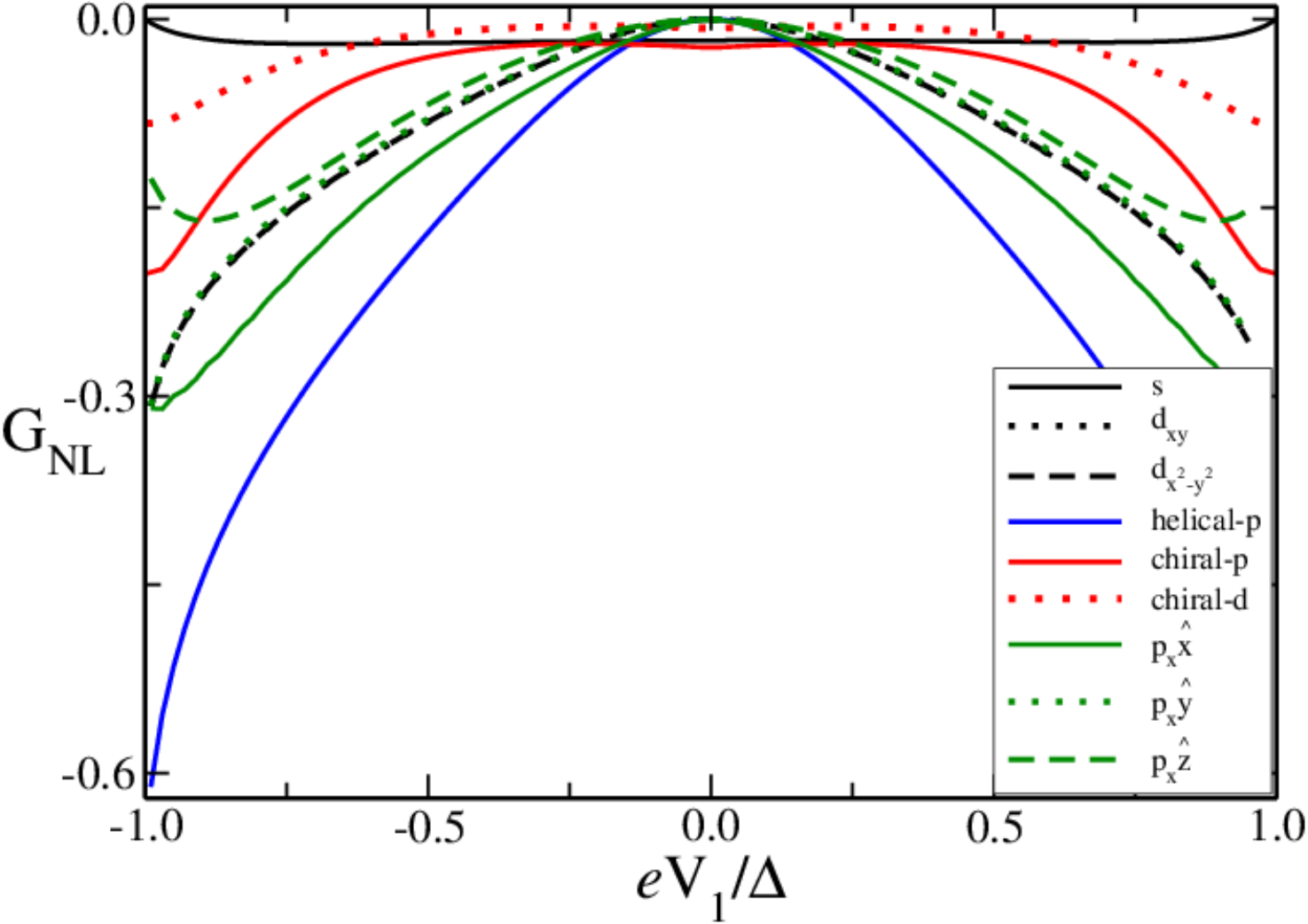}
\caption{Differential non-local conductance ($G_{NL}$, see Eq.~\ref{eqn:G}) in units of $2e^{2}/h$, for $N_1$/I/US/I/$N_2$ junction vs bias voltage ($e$V$_1/\Delta$) for US with different pairing symmetries where $k_F R=55$, $R / \xi =2$, $z=1$ and $eV_2/ \Delta=0$. }
\label{GNLv}
\end{figure}
In Fig.~\ref{GNLv} we plot differential non-local conductance G$_{NL}$ vs. bias voltage ($eV_1$) with $eV_2/ \Delta=0.0$ for intermediate barrier strength ($z=1$). $G_{NL}$, dominated by EC is fully negative for entire range of bias voltage($-1 < eV_1/ \Delta < 1$) irrespective of change in pairing potential for different pairing symmetries. In Appendix A, we describe the crossed Andreev conductance ($G_{CAR}$) and elastic co-tunneling ($G_{EC}$) contribution to differential non-local conductance for each pairing symmetry in the non-local setup. \\
Theoretically, it has been shown that for transparent limit, non-local conductance~\cite{gnl,floser} is negative, which does not convey enough information about Cooper pair splitting, known as CAR. It motivates us to study shot noise cross-correlations to differentiate between different pairing symmetries.

\begin{figure}[h!]
\includegraphics[scale=0.32]{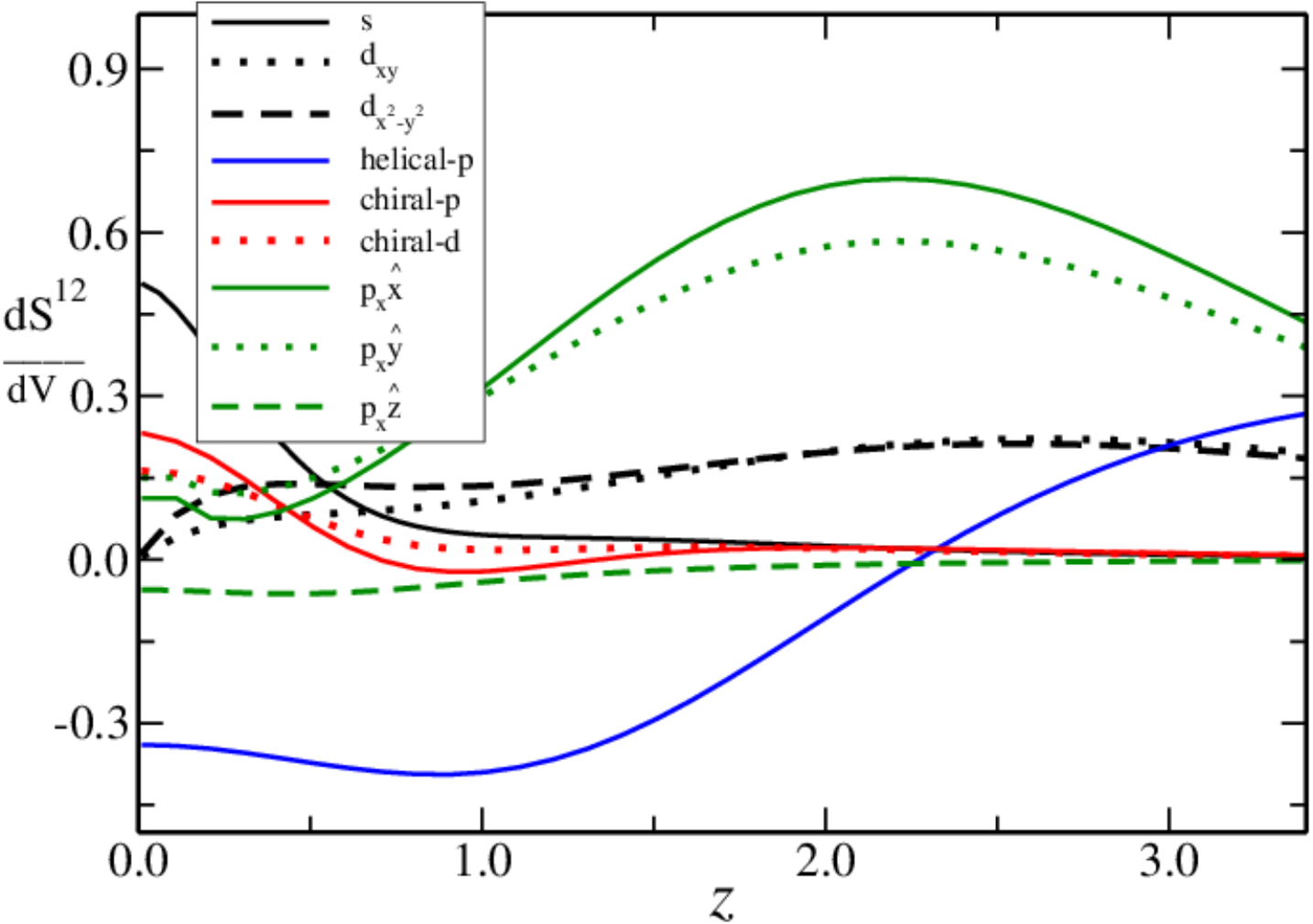}
\caption{Differential shot noise cross-correlations (symmetric setup, see Eq.~\ref{eqn:dsn sym}) in units of $4 |e|^3/h$ for $N_1$/I/US/I/$N_2$ junction vs barrier strength ($z$), with different pairing symmetries where $k_F R=55$, $R / \xi =2 $, $eV_1/ \Delta =$ $eV_2/ \Delta=0.2$.}
\label{dsnz symm}
\end{figure}

Next, let us first look at differential shot noise cross-correlation behavior for different pairing symmetries due to changes in barrier strength. First, for symmetric setup ($eV_1 / \Delta=eV_2 / \Delta=0.2$) as shown in Fig.~\ref{dsnz symm}. Differential shot noise cross-correlations ($dS^{12}/dV$) for helical-$p$ pairing changes sign with increase in barrier strength ($z$). Positive $dS^{12}/dV$ in case helical-$p$ pairing indicates Cooper pair splitting seen in tunneling regime ($z > 2$), which we explain in detail in the next section. $dS^{12}/dV$ is enhanced in the fully transparent ($z\rightarrow 0$) limit for gapful non-topological $s$-wave and gapful topological chiral pairings, while it vanishes in tunneling regime ($z \rightarrow $large). In the transparent limit, $dS^{12}/dV$ slowly increases with increase in barrier strength for non-topological nodal singlet pairing ($d_{xy}$ and $d_{x^2-y^2}$). Increase in barrier strength enhances $dS^{12}/dV$ for topological nodal triplet pairing ($p_{x} \hat{x}$ and $p_{x} \hat{y}$). $dS^{12}/dV$ is negative in the transparent limit for non-topological nodal triplet pairing ($p_{x} \hat{z}$) and vanishes in tunneling regime.
\begin{figure}[h!]
\hspace{-0.15cm}
\includegraphics[scale=0.32]{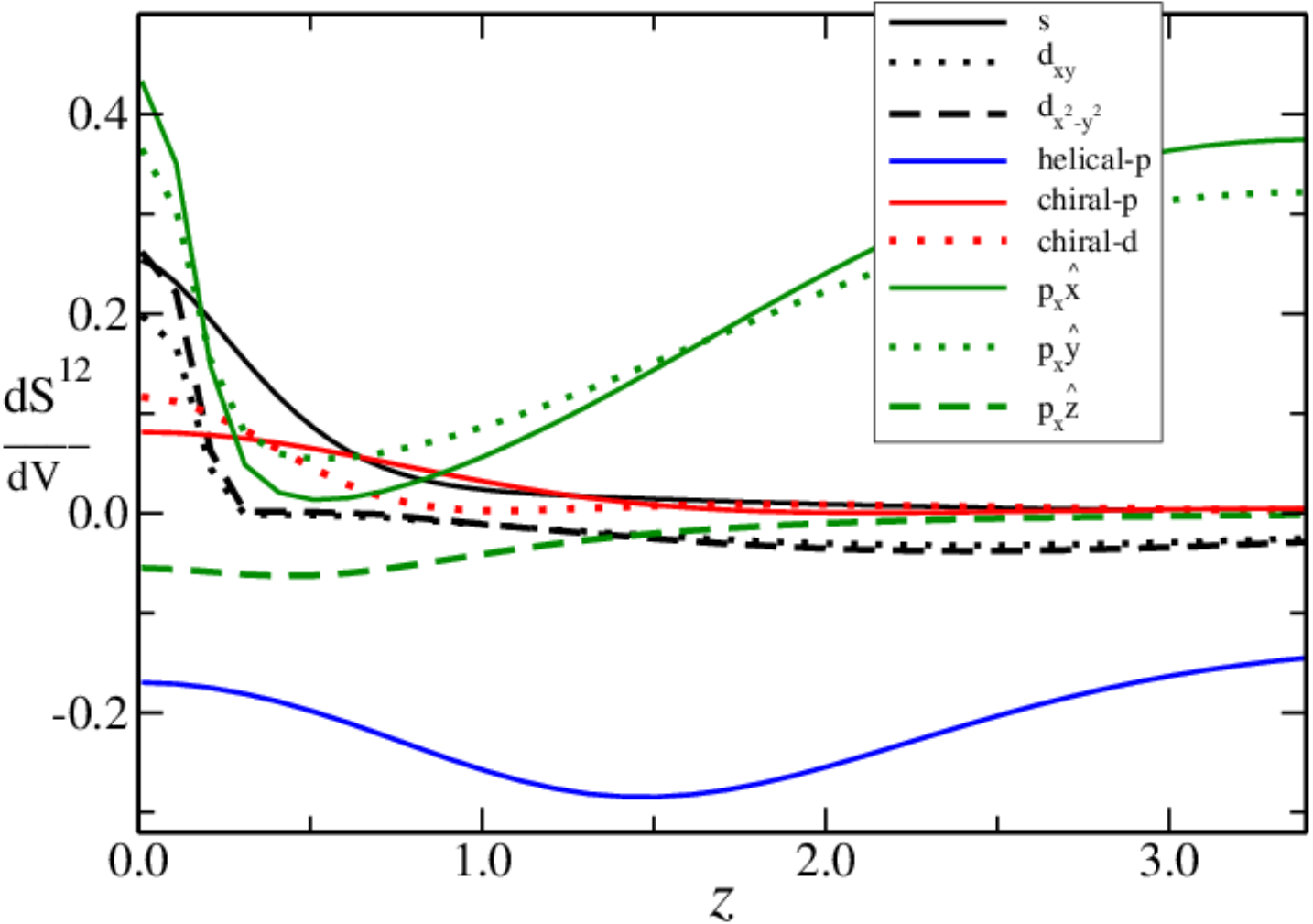}
\caption{Differential shot noise cross-correlations (non-local setup, see Eq.~\ref{eqn:dsn asym}) in units of $4 |e|^3/h$ for $N_1$/I/US/I/$N_2$ junction vs barrier strength ($z$) for different pairing symmetries of the unconventional superconductor, with $k_F R=55$, $R / \xi =2 $, $eV_1/ \Delta =0.2$ and $eV_2 / \Delta=0$ . }
\label{dsnz asym}
\end{figure}

In Fig.~\ref{dsnz asym} we plot $dS^{12}/dV$ vs $z$ for non-local set up with $eV_1/ \Delta=0.2$ and $eV_2/ \Delta=0.0$. Similar results as symmetric setup are obtained for gapful $s$, chiral pairing, and nodal $p_x \hat{z}$ pairings in non-local setup. Unlike symmetric setup, $dS^{12}/dV$ is enhanced in the case of non-local setup for topological nodal pairings in the transparent limit. $dS^{12}/dV$ for helical-$p$ pairing changes from negative to positive as one goes from transparent to tunneling regime in symmetric setup, while for non-local setup, it is completely negative. The opposite behavior is seen for $d_{xy}$ and $d_{x^2-y^2}$ pairing as $dS^{12}/dV$ is completely positive for symmetric setup. However, it changes from positive to negative as one goes from transparent to tunneling regime in a non-local setup. For other cases, $dS^{12}/dV$ is always positive for both symmetric and non-local setups. Differential shot noise cross-correlations can be a good indicator for helical-$p$, $p_x \hat{z}$, $d_{xy}$ and $d_{x^2-y^2}$ pairings.
\begin{figure}[h!]
\hspace{-0.5cm}
\includegraphics[scale=0.32]{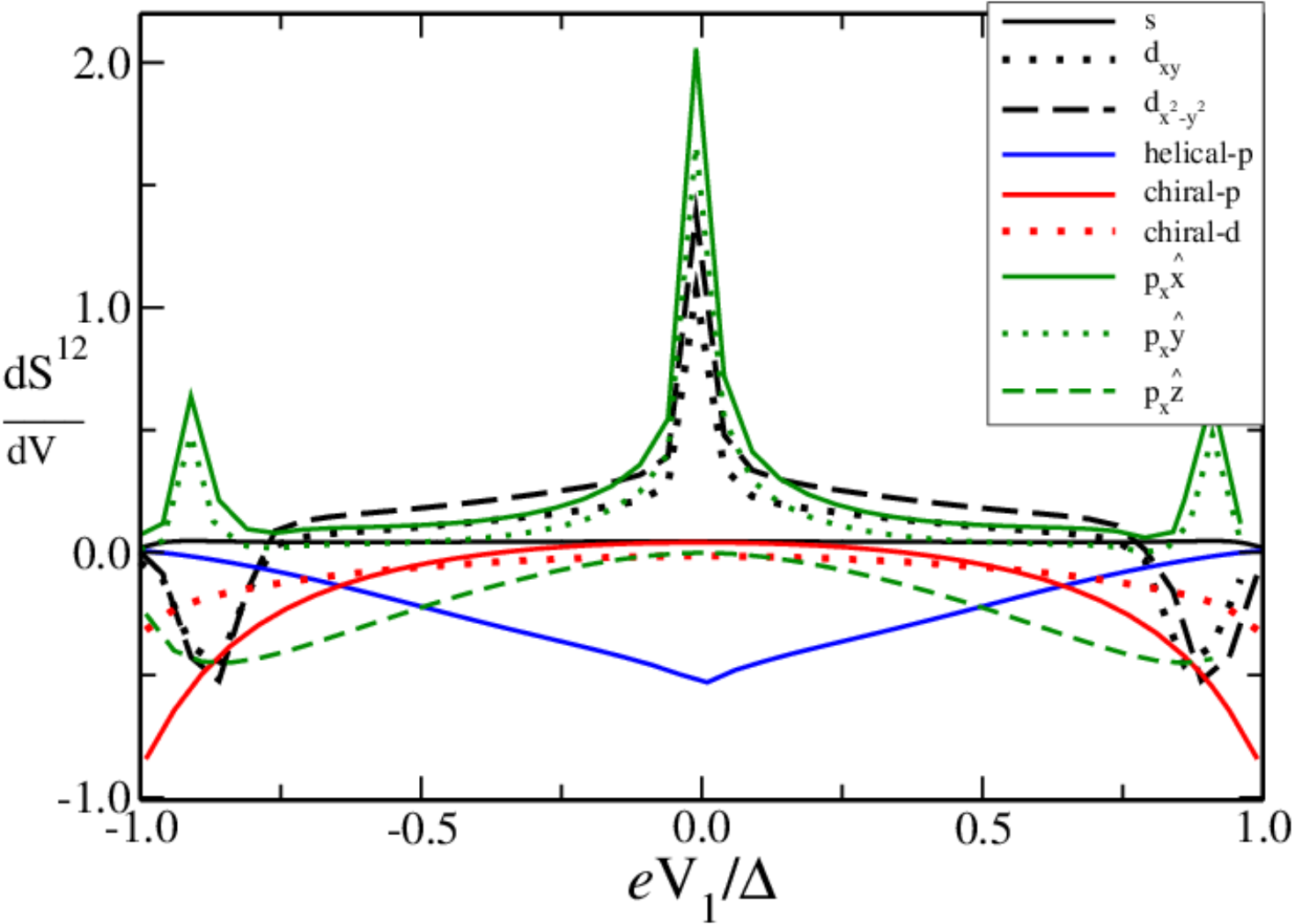}
\caption{Differential shot noise cross-correlations (for symmetric setup) in units of $4 |e|^3/h$ for $N_1$/I/US/I/$N_2$ junction vs bias voltage ($e$V$_1/\Delta$) for different pairing symmetries of the unconventional superconductor, with $k_F R=55$, $R / \xi =2$, $eV_1=eV_2$ and $z=1$.}
\label{dsnv symm}
\end{figure}

Next, in Fig.~\ref{dsnv symm} we plot $dS^{12}/dV$ for different pairing symmetries as a function of bias voltage ($eV_1/\Delta$) in symmetric set up for intermediate barrier strength ($z=1$). $dS^{12}/dV$ shows zero bias peak (ZBP) for both nodal non-topological $d_{xy}$ and $d_{x^2-y^2}$ and nodal topological $p_x \hat{x}$ and $p_x \hat{y}$ pairings. $dS^{12}/dV$ for gapful topological pairings (helical-$p$, chiral-$p$ and chiral-$d$) and non-topological triplet $p_x \hat{z}$ pairing are negative for entire range of bias voltage. $dS^{12}/dV$ shows a zero bias dip (ZBD) for helical-$p$, while it vanishes for chiral-$p$ pairing at zero bias. When bias voltages tend to superconducting gap $\pm \Delta$, $dS^{12}/dV$ is strongly enhanced for chiral pairings.
\begin{figure}[h!]
\includegraphics[scale=0.3]{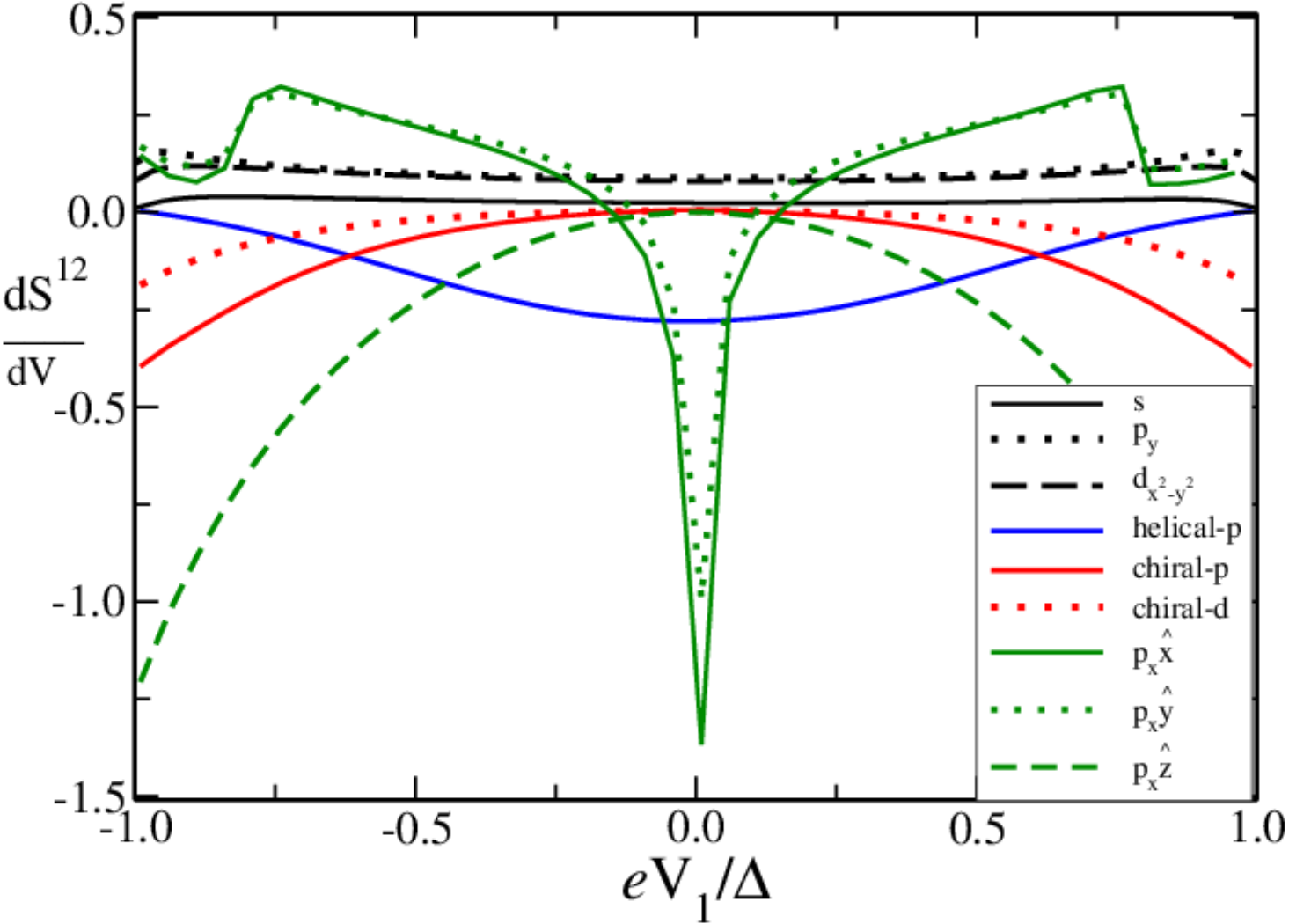}
\caption{Differential shot noise cross-correlations (for non-local setup) in units of $4 |e|^3/h$ for $N_1$/I/US/I/$N_2$ junction vs bias voltage ($eV_1/\Delta$) for different pairing symmetries with $k_F R=55$, $R / \xi =2 $, $eV_2 / \Delta=0$ and $z=1$.}
\label{dsnv asym}
\end{figure}

Next, we plot $dS^{12}/dV$ vs bias voltage ($eV_1/\Delta$) for non-local set up with $eV_2/\Delta=0.0$ in Fig.~\ref{dsnv asym}. Tuning the bias voltage $eV_2/\Delta$ to zero does not affect $dS^{12}/dV$ for gapful topological pairing (chiral-$p$, chiral-$d$ and helical-$p$) which show similar behaviour as in the symmetric setup, shown in Fig.~\ref{dsnv symm}. Contrary to symmetric setup, in case of non-local setup for nodal topological pairing ($p_x \hat{x}$, $p_x \hat{y}$) $dS^{12}/dV$ shows ZBD instead of ZBP. Table~\ref{table dsn} summarizes the results for non-local conductance and differential shot noise (in both symmetric and non-local setups). Table~\ref{table dsn} succinctly puts all results in perspective. In the next subsection, we plot HBT or shot noise cross-correlations in the tunneling and transparent regimes.

\subsection{Shot noise cross-correlations}
\begin{figure}[h!]
\includegraphics[scale=0.3]{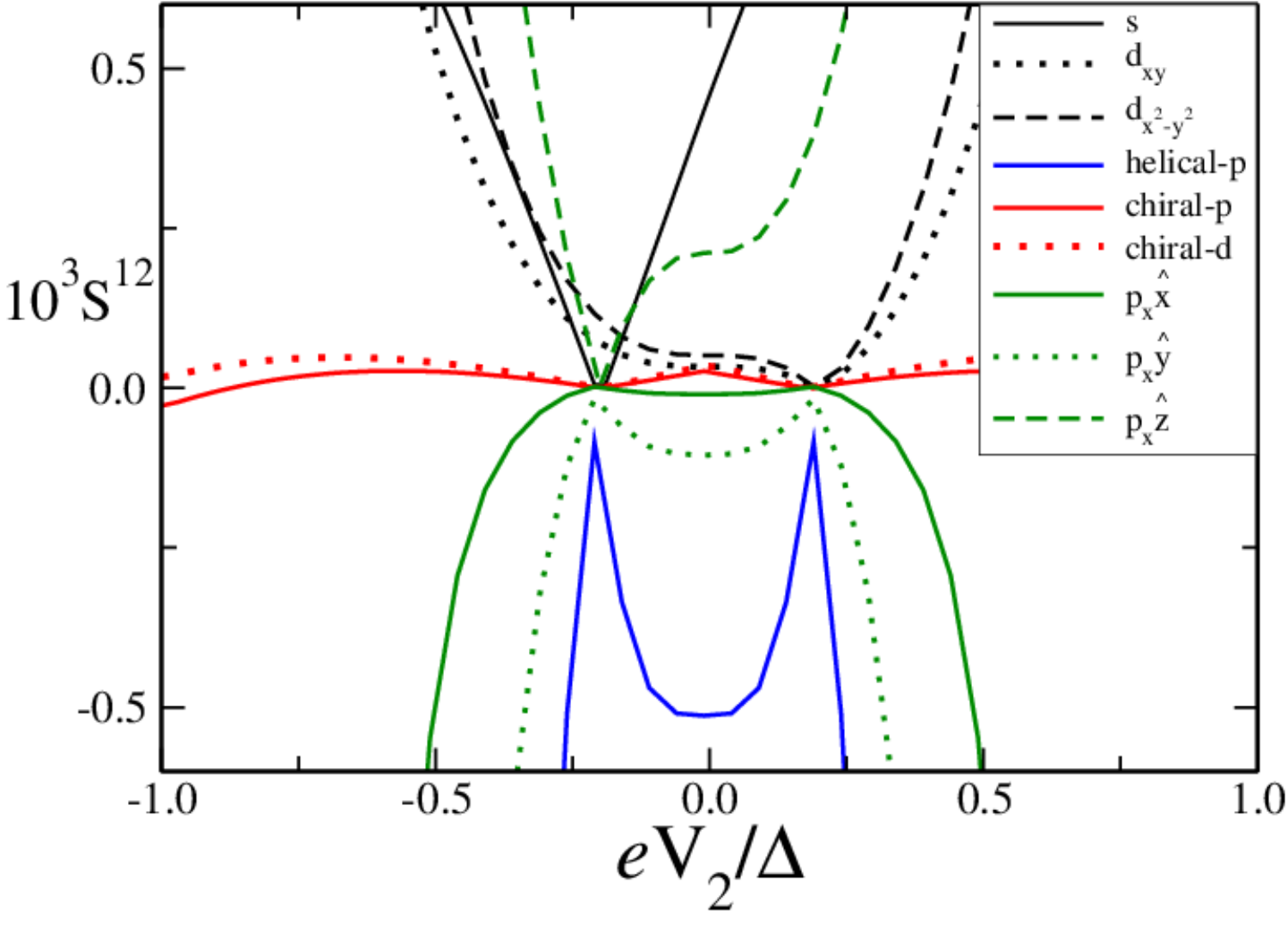}
\caption{Shot noise cross-correlations ($S^{12}$ multiplied by $10^{3}$) in units of $4 |e|^2/h$ for $N_1$/I/US/I/$N_2$ junction vs bias voltage ($eV_2/\Delta$) for different pairing symmetries with $k_F R=55$, $R / \xi =2 $, $eV_1 / \Delta=0.2$ and $z=0.1$ (Transparent barriers).}
\label{snv trans}
\end{figure}

\begin{figure}[h!]
\includegraphics[scale=0.3]{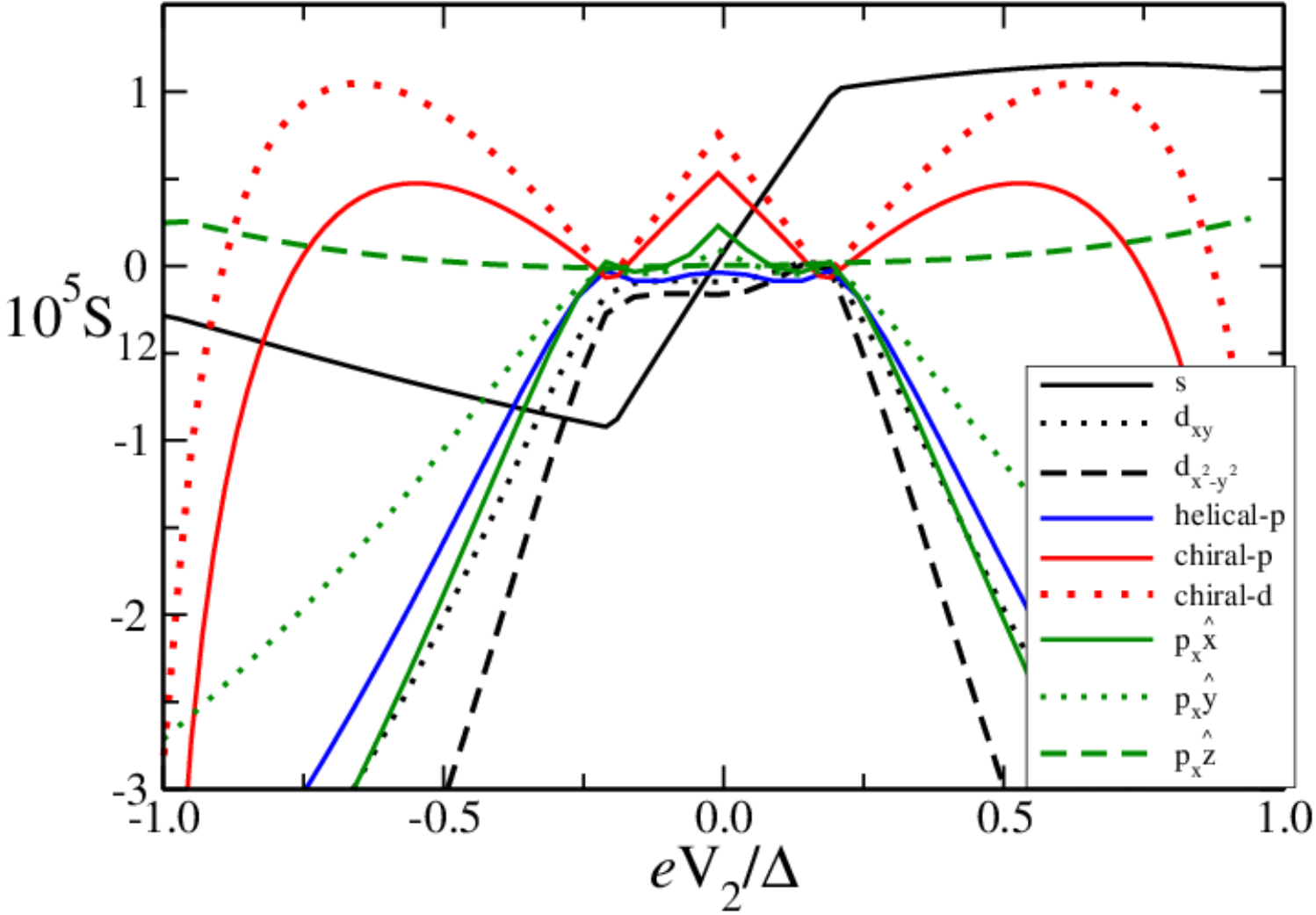}
\caption{Shot noise cross-correlations ($S^{12}$ multiplied by $10^{5}$) in units of $4 |e|^2/h$ for $N_1$/I/US/I/$N_2$ junction vs bias voltage ($eV_2/\Delta$) for different pairing symmetries with $k_F R=55$, $R / \xi =2 $, $eV_1 / \Delta=0.2$ and $z=3$ (Tunnel barriers).}
\label{snv tun}
\end{figure}

Shot noise cross-correlations $S^{12}$ for the setup (Fig.~\ref{fig:NUSN}) in the general case (neither symmetric nor non-local) is calculated using Eq.~\ref{eqn:sn} and plotted as function of bias voltage $eV_2/ \Delta$ applied to normal metal $N_2$ for both transparent ($z=0.1$) junction, see Fig.~\ref{snv trans}, and for tunnel limit ($z=3$) in Fig.~\ref{snv tun}. Shot noise cross-correlations for $s$-wave comes in line with previous results~\cite{gcar} for both tunnel and transparent limit.

In Fig.~\ref{snv trans}, for superconductors that do not possess spin rotation symmetry, i.e., for gapful helical-$p$, nodal $p_x \hat{x}$ and $p_{x} \hat{y}$ cases, shot noise cross-correlations ($S^{12}$) are negative with a dip at zero bias in the transparent limit. Positive HBT correlations are seen in the transparent limit for superconductors with spin rotation symmetry. $S^{12}$ for topological chiral-$p$, chiral-$d$ pairings shows a ZBP and is symmetric as function of bias voltage whereas $S^{12}$ for all non-topological pairings, i.e., $s$-wave, $p_x \hat{z}$, $d_{xy}$ and $d_{x^2-y^2}$, show asymmetric behaviour as function of bias voltage $V_2$, vanishing at $V_2=-V_1$ for $s$-wave and $p_x \hat{z}$ and at $V_2=V_1$ for $d_{xy}$ and $d_{x^2-y^2}$ pairings. Hence, by tuning bias voltages, we can control transport; for example, in the case of topological pairings, there is no transport for both $V_2=\pm V_1$ whereas
for non-topological pairings either at $V_2=-V_1$ or at $V_2=V_1$.

In the tunnel limit (Fig.~\ref{snv tun}), positive shot noise cross-correlations for chiral-$p$ and chiral-$d$ pairings show similar behavior with a ZBP as in the transparent limit. However, contrary to the transparent limit, $S^{12}$ for helical-$p$ pairing in the tunnel limit vanishes at zero bias. Shot noise cross-correlations for $s$-wave pairing linearly increase as a function of $V_2$. Similar to transparent limit, in tunnel limit $S^{12}$ for all non-topological cases are asymmetric as function of bias voltage and vanish a particular one bias voltage (either $V_2=V_1$ or $V_2=-V_1$), whereas for all topological cases $S^{12}$ vanishes at both $V_2=V_1$ and $V_2=-V_1$, and is symmetric. Negative shot noise cross-correlations are seen for nodal non-topological singlet ($d_{xy}$, $d_{x^2-y^2}$) pairings which vanish at $V_2=V_1$. At zero bias, $S^{12}$ shows ZBP for topological $p_x \hat{x}$ and $p_x \hat{y}$ whereas it is flat at zero bias for non-topological $p_x \hat{z}$, enabling a distinction between the topological and non-topological $p_x$ pairing. Table~\ref{table sn} summarizes the results for shot noise cross-correlations in both transparent and tunnel limits.

\subsection{Processes in play}
Shot noise cross-correlations have been calculated in metal/superconductor/metal hybrid junctions to study Cooper pair splitting; see Refs.~\cite{NSN,bignon,floser}. Shot noise cross-correlations for $s$-wave superconductor show linear behaviour, which is $\propto V_2$ in tunnel limit and $\propto eV_1 + eV_2$ in transparent limit for bias voltage range ($-V_1 < V_2 <V_1$) as seen in Figs.~\ref{snv trans} and \ref{snv tun}. This has been also predicted in Refs.~\cite{NSN,bignon,golubev,arXiv:2101} and explained in Ref.~\cite{floser}. Shot noise cross-correlations for $s$-wave pairing vanish at $V_2=-V_1$ in transparent limit but vanish at $V_2=0$ in tunnel limit, which was also predicted in Ref.~\cite{golubev}. One understands this behavior by dividing the shot noise correlations into individual contributions to shot noise from local (either Andreev reflection or normal reflection) processes and non-local (CAR or EC) processes. In the subsections below, we try to understand the reasons for the plots shown in Fig.~\ref{snv trans} (transparent limit) and Fig.~\ref{snv tun} (tunnel limit) for HBT correlations via these processes for all pairing symmetries.

\subsubsection{Tunnel limit}
Shot noise cross-correlations, from Eq.~(\ref{eqn:snheaviside}) of Appendix, consists of local (AR-Andreev reflection, NR-normal reflection) amplitudes and non-local (CAR- Crossed Andreev reflection, EC-elastic cotunneling) amplitudes. Each term in the shot noise cross-correlations Eq.~(\ref{eqn:snheaviside}) consists of four processes which can be grouped as EC-NR, CAR-NR, EC-AR, CAR-AR, and a mixed group of all four processes. EC-NR implies product of elastic cotunneling and normal reflection amplitudes, such as $s^{ee}_{12} s^{ee}_{21} s^{ee *}_{22} s^{ee*}_{11}$, similarly CAR-NR is product of crossed Andreev reflection and normal reflection amplitudes, such as $s^{eh}_{21} s^{he}_{12} s^{hh *}_{11} s^{ee *}_{22}$, CAR-AR is product of crossed Andreev reflection and Andreev reflection amplitudes, such as $s^{he}_{21} s^{he}_{12} s^{he *}_{11} s^{he *}_{22}$ and EC-AR is product of elastic cotunneling and Andreev reflection amplitudes, such as $s^{he}_{21} s^{he}_{12} s^{he *}_{11} s^{he *}_{22}$.

Inspecting the different contributions for $s$-wave case, we see that NR amplitudes ($s^{hh}_{11}, s^{ee}_{22}$) $\rightarrow 1$ in tunnel limit, i.e., large $z$, hence, CAR-NR terms in $S^{12}$, see Eq.~(\ref{eqn:snheaviside}), reduce to just CAR (or $s_{CAR}=s^{eh}_{21} s^{he}_{12}$) and EC-NR reduces to just EC, (or $s_{EC}=s^{ee}_{12} s^{ee}_{21}$). EC contribution to shot noise cross-correlations in tunnel limit, using electron-hole symmetry of scattering matrix amplitudes gives $S^{EC}=s_{EC} (h_1+h_2)$, here $h_1,h_2$ are Heaviside theta functions and given in Eq.~(\ref{eqn:heaviside}), while CAR contribution to shot noise cross-correlations in tunnel limit, which again using electron-hole symmetry gives $S^{CAR}= s_{CAR} (h_3+ h_4 )$, where $h_3,h_4$ are given in Eq.~(\ref{eqn:heaviside}). Shot noise cross-correlations for $s$-wave case in tunnel limit in small bias voltage regime ($-V_1< V_2<V_1$) can be written as $S^{12} = S^{CAR} + S^{EC}$, with $S^{EC}= s^{ee}_{12} s^{ee}_{21} (\Theta (e|V_1 - E) - \Theta (|e|V_2 - E)- \Theta (-|e|V_1 - E) + \Theta (-|e|V_2 - E))$ and $S^{CAR}= s^{eh}_{21} s^{he}_{12} (-\Theta (e|V_1 - E) - \Theta (|e|V_2 - E)+ \Theta (-|e|V_1 - E) + \Theta (-|e|V_2 - E))$, where $\Theta$ is Heaviside theta function. However, for other pairing symmetries, NR amplitudes ($s^{hh}_{11}, s^{ee}_{22}$) $\centernot \rightarrow 1$ in tunnel limit. Thus, $S^{CAR-NR}$ does not reduce to $S^{CAR}$, and $S^{EC-NR}$ does not reduce to $S^{EC}$ for these cases. For example, in case of chiral-$p$, chiral-$d$ and $p_x \hat{z}$ pairings, $S^{12} = S^{CAR-NR}+ S^{EC-NR}$, where $S^{CAR-NR} = (s^{eh}_{21} s^{he}_{12} s^{hh *}_{11} s^{ee *}_{22}) (h_3+ h_4)$ and $S^{EC-NR} = (s^{ee}_{12} s^{ee}_{21} s^{ee *}_{22} s^{ee*}_{11})( h_1+ h_2)$.

\begin{widetext}

\begin{figure}[h!]
\includegraphics[scale=1.8]{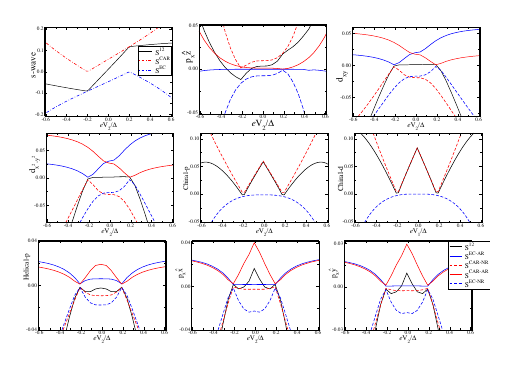}
\caption{Processes contributing to shot noise cross-correlations in units of $(10^4)$ $4|e|^2/h$ for $N_1$/I/US/I/$N_2$ junction vs bias voltage ($eV_2/\Delta$) for non-topological and topological superconductors with $k_F R=55$, $R / \xi =2 $, $eV_1 / \Delta=0.2$, $z=3$ (tunnel barriers).}
\label{sn tunnel z3}
\end{figure}

\end{widetext}

From Fig.~\ref{sn tunnel z3}, unlike $s$-wave (non-topological gapful), shot noise cross-correlations for chiral-$p$ and chiral-$d$ (topological gapful) pairings in tunnel limit are exclusively due to the CAR-NR process at low bias voltages as EC-NR is suppressed in this regime.

In Fig.~\ref{sn tunnel z3}, essential processes contributing to $S^{12}$ are CAR-NR, CAR-AR, EC-NR, and EC-AR, which vanish at $V_2=V_1$ or $V_2=-V_1$ or both $V_2=\pm V_1$ resulting in no transport at these values for all cases. For chiral (both $p$ and $d$) superconductors, CAR-NR and EC-NR contribute to HBT correlations. However, EC-NR contribution to HBT correlations vanish at low bias voltages but dominates at $eV_2 \rightarrow \Delta$, which leads to a change in sign from positive to negative of shot noise cross-correlations. Flat $S^{12}$ is seen at zero bias for $p_x \hat{z}$ pairing due to the contribution of both EC-NR and CAR-NR.

It has been before predicted for $s$-wave pairing, in Ref.~\cite{bignon}, shot noise cross-correlations in our calculation for non-topological $s$-wave superconductor in the tunnel limit in low bias voltage range $(-V_1<V_2 <V_1)$ can be written as :
\begin{eqnarray}
S^{12}_{tunnel}(s) \propto eV_2.
\label{eqn:sn tunnel}
\end{eqnarray}

Similarly $S^{12}$, in tunnel limit, for gapful topological (chiral-$p$, chiral-$d$, helical-$p$), nodal topological ($p_x \hat{x}$, $p_x \hat{y}$), nodal non-topological superconductors ($p_x \hat{z}$, $d_{xy}$, $d_{x^2-y^2}$) in low bias voltage range $(-V_1<V_2 <V_1)$, one gets :
\begin{eqnarray}
S^{12}_{tunnel} (p_{x} \hat{z}) \propto (eV_2+ eV_1)^2-c, \nonumber \\
S^{12}_{tunnel} (d_{xy},~d_{x^2-y^2}) \propto -(c~eV_2- eV_1)^2, \nonumber \\
S^{12}_{tunnel} (\text{chiral}-p\text{ and chiral}-d) \propto (eV_1-|eV_2|), \nonumber \\
S^{12}_{tunnel} (\text{helical}-p) \propto -(eV_1~|eV_2|-(eV_2)^2), \nonumber \\
S^{12}_{tunnel} (p_x \hat{x},~p_x \hat{y}) \propto (-|eV_2|+c~eV_1)-|(eV_2)^2-(eV_1)^2|,
\label{eqn:sn tunnel chiral}
\end{eqnarray}
where $c$ is a constant term. For topological superconductors, $S^{12}$ in tunnel limit is symmetric to change in sign of bias voltage ($V_2$), whereas for non-topological superconductors, it is asymmetric as a function of bias voltage ($V_2$). This symmetry can be a marker also for the presence of Majorana fermions akin to ZBCP in metal/topological superconductor junction. Shot noise cross-correlations predicted in our work for $s$-wave are in line with that seen for $s$ wave superconductor in Refs.~\cite{bignon,golubev}.

\begin{widetext}

\begin{table}[h!]
\centering
\caption{Processes that contribute to shot noise cross-correlations are denoted by $\checkmark$ and $\times$ represents the processes that do not contribute to shot noise cross-correlations at a low bias voltage range ($-V_1<V_2< V_1$) for each pairing in tunnel limit.}
\scalebox{0.72}{\begin{tabular}{|c|c|c|c|c|c|c|}
\hline
\begin{tabular}[c]{@{}c@{}}\\\\\\Topology\end{tabular} & Type & Pairing & \begin{tabular}[c]{@{}c@{}}EC-AR\\ ($\langle \Delta I^e_{N_1} \Delta I^h_{N_2} \rangle + \langle \Delta I^h_{N_1} \Delta I^e_{N_2} \rangle$)\end{tabular} & \begin{tabular}[c]{@{}c@{}}CAR-AR\\ ($\langle \Delta I^e_{N_1} \Delta I^e_{N_2} \rangle + \langle \Delta I^h_{N_1} \Delta I^h_{N_2} \rangle$)\end{tabular} & \begin{tabular}[c]{@{}c@{}}EC-NR\\ ($\langle \Delta I^e_{N_1} \Delta I^e_{N_2} \rangle + \langle \Delta I^h_{N_1} \Delta I^h_{N_2} \rangle$)\end{tabular} & \begin{tabular}[c]{@{}c@{}}CAR-NR\\ ($\langle \Delta I^e_{N_1} \Delta I^h_{N_2} \rangle + \langle \Delta I^h_{N_1} \Delta I^e_{N_2} \rangle$)\end{tabular} \\
\hline
\multirow{3}{*}{\begin{tabular}[c]{@{}c@{}}Non\\Topological\end{tabular}} & Gapful & $s$ & $\times$ & $\times$ & $\checkmark$ & $\checkmark$ \\
\cline{2-7}
& \multirow{2}{*}{Nodal} & $p_x \hat{z}$ & $\times$ & $\times$ & $\checkmark$ & $\checkmark$ \\
\cline{3-7}
& & \begin{tabular}[c]{@{}c@{}}$d_{x^2-y^2},$ \\ $d_{xy}$\end{tabular} & $\checkmark$ & $\checkmark$ & $\checkmark$ & $\checkmark$ \\
\hline
\multirow{3}{*}{Topological} & \begin{tabular}[c]{@{}c@{}}Chiral\\(Gapful)\end{tabular} & \begin{tabular}[c]{@{}c@{}}$p_x+i p_y,$ \\~$d_{x^2-y^2}+ i d_{xy} $\end{tabular} & $\times$ & $\times$ & $\checkmark$ & $\times$ \\
\cline{2-7}
& \begin{tabular}[c]{@{}c@{}}Helical\\(Gapful)\end{tabular} & $p$ & $\checkmark$ & $\checkmark$ & $\checkmark$ & $\checkmark$ \\
\cline{2-7}
& Nodal & \begin{tabular}[c]{@{}c@{}}$p_x\hat{x},$~ \\$p_x \hat{y}$\end{tabular} & $\checkmark$ & $\checkmark$ & $\checkmark$ & $\checkmark$ \\
\hline
\end{tabular}}
\label{pplay tun}
\end{table}

\end{widetext}

Table~\ref{pplay tun} summarizes how the different processes contribute to the shot noise cross-correlations as a function of bias voltages. Current is carried by electrons or holes, i.e., $I^e$ or $I^h$. Shot noise cross-correlations can be categorized based on correlation between same type of carriers, i.e., $\langle \Delta I^e_{N_1} \Delta I^e_{N_2} \rangle + \langle \Delta I^h_{N_1} \Delta I^h_{N_2} \rangle$ for EC-NR and CAR-AR or between different types of carriers, i.e., $\langle \Delta I^e_{N_1} \Delta I^h_{N_2} \rangle + \langle \Delta I^h_{N_1} \Delta I^e_{N_2} \rangle$ for EC-AR and CAR-NR. Hence, EC-AR and CAR-NR behave similarly due to different charge carrier correlations, e.g., in the case of non-topological $d_{xy}$ and $d_{x^2-y^2}$ pairings, while EC-NR and CAR-AR behave similarly due to correlations between same charge carriers, e.g., in case of helical-$p$, $p_x \hat{x}$ and $p_x \hat{y}$ pairings which are topological. For non-topological $p_x \hat{z}$ pairing and topological chiral-$p$ and chiral-$d$ pairing, AR contribution tends to be negligible, hence, suppressing EC-AR and CAR-AR contributions at low bias voltages.

\subsubsection{Transparent limit}
From Ref.~\cite{floser}, for $s$-wave pairing, shot noise cross-correlations in transparent limit contribute only from the EC-AR process. Along with $s$-wave, shot noise cross-correlations for chiral-$p$, chiral-$d$ and $p_x \hat{z}$ pairings in transparent limit ($z=0.0$) are limited to EC-AR, as CAR-NR contribution is negligible. Thus, for $s$-wave, chiral-$p$, chiral-$d$ and $p_x \hat{z}$ pairings, $S^{12}=S^{EC-AR}= s_{EC-AR} h_{EC-AR}$, where $s_{EC-AR} = s^{ee}_{12} s^{hh}_{21} s^{eh *}_{11} s^{he *}_{22}$ and $h_{EC-AR} = h_3 + h_4$, $h_3$ and $h_4$ are Heaviside theta functions given in Eq.~(\ref{eqn:heaviside}). Simplifying the shot noise cross-correlations for $s$-wave, chiral-$p$, chiral-$d$ and $p_x \hat{z}$ pairings in transparent limit ($z=0.0$), and in low bias voltage regime $-V_1< V_2<V_1$, we get $S^{12}=S^{EC-AR}= (s^{ee}_{12} s^{hh}_{21} s^{eh *}_{11} s^{he *}_{22}) (-\Theta (e|V_1 - E) - \Theta (|e|V_2 - E)+ \Theta (-|e|V_1 - E) + \Theta (-|e|V_2 - E))$, where $\Theta$ is Heaviside theta function. All four processes (EC-AR, EC-NR, CAR-AR, and CAR-NR) contribute to shot noise cross-correlations for superconductors that do not possess spin rotation symmetry. In contrast, only one process contributes to shot noise cross-correlations for superconductors that possess spin rotation symmetry in a low bias voltage range. Hence, for $d_{xy}$ and $d_{x^2-y^2}$ cases, $S^{12}=S^{CAR-AR}$ at low bias voltages. CAR-AR contribution to shot noise can be written as $S^{CAR-AR}= s_{CAR-AR} h_{CAR-AR}$, where $s_{CAR-AR} = s^{eh}_{21} s^{he}_{12} s^{he *}_{11} s^{eh *}_{22}$ and $h_{CAR-AR} = h_1 + h_4$, $h_1$ and $h_4$ are Heaviside theta functions given in Eq.~\ref{eqn:heaviside}.

\begin{widetext}

\begin{figure}[h!]
\includegraphics[scale=1.8]{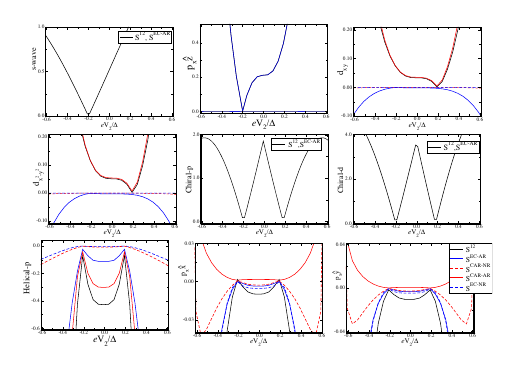}
\caption{Processes contributing to shot noise cross-correlations in units of $4|e|^2/h$ for $N_1$/I/US/I/$N_2$ junction vs bias voltage ($eV_2/\Delta$) for non-topological superconductors $S^{12}(10^3)$ and topological superconductors $S^{12}(10^5)$ with $k_F R=55$, $R / \xi =2 $, $eV_1 / \Delta=0.2$ and $z=0.0$ (transparent limit).}
\label{sn transparent0}
\end{figure}

\end{widetext}

Similar to tunnel limit in the transparent limit, processes that contribute to $S^{12}$ are asymmetric as function of $V_2$ and lead to no transport at $V_2=-V_1$ for $s$-wave case and at $V_2=V_1$ for $d_{xy}$ and $d_{x^2-y^2}$ cases as given in Eq.~\ref{eqn:sn trans}. For all topological cases, processes that contribute to $S^{12}$ are symmetric as function of $V_2$ resulting in vanishing $S^{12}$ at both $V_2= V_1$ and $V_2=- V_1$ as shown in Fig.~\ref{sn transparent0}. Shot noise cross-correlations for all pairing symmetries in Fig.~\ref{sn transparent0} shown for $z=0.0$ (transparent limit) show similar behaviour as also seen in Fig.~\ref{snv trans} for $z=0.1$.

As also predicted in Ref.~\cite{golubev}, shot noise cross-correlations for $s$-wave superconductor in transparent limit at zero temperature reduces to,
\begin{eqnarray}
S^{12}_{transparent} (s) \propto eV_1 + eV_2.
\label{eqn:sn trans}
\end{eqnarray}

In the transparent limit and low bias voltage range $(-V_1<V_2 <V_1)$, shot noise cross-correlations for the rest of the pairing symmetries are given as,
\begin{eqnarray}
S^{12}_{transparent} (\text{chiral}-p, \text{chiral}-d) \propto eV_1- |eV_2|, \nonumber \\
S^{12}_{transparent} (p_{x} \hat{z}) \propto (c~eV_2+ eV_1)^2, \nonumber \\
S^{12}_{transparent} (d_{xy},~d_{x^2-y^2}) \propto (c~eV_2- eV_1)^2, \nonumber \\
S^{12}_{transparent}(\text{helical $p$, $p_x \hat{x}$, $p_x \hat{y}$}) \propto (eV_2)^2-(eV_1)^2,
\label{eqn:sn trans topo}
\end{eqnarray}
where $c$ is a constant value. Bias voltage dependence of shot noise cross-correlations for chiral pairing is the same in both transparent and tunnel limits. In contrast, barrier strength influences bias voltage dependence of shot noise cross-correlations for other cases.

\begin{widetext}

\begin{table}[h!]
\centering
\caption{Processes that contribute to shot noise cross-correlations are denoted by $\checkmark$ and $\times$ represents the processes that do not contribute to shot noise cross-correlations at low bias voltage ($-V_1<V_2<V_1$) for each pairing in transparent limit.}
\scalebox{0.72}{\begin{tabular}{|c|c|c|c|c|c|c|}
\hline
\begin{tabular}[c]{@{}c@{}}\\\\\\Topology\end{tabular} & Type & Pairing & \begin{tabular}[c]{@{}c@{}}EC-AR\\ ($\langle \Delta I^e_{N_1} \Delta I^h_{N_2} \rangle + \langle \Delta I^h_{N_1} \Delta I^e_{N_2} \rangle$)\end{tabular} & \begin{tabular}[c]{@{}c@{}}CAR-AR\\ ($\langle \Delta I^e_{N_1} \Delta I^e_{N_2} \rangle + \langle \Delta I^h_{N_1} \Delta I^h_{N_2} \rangle$)\end{tabular} & \begin{tabular}[c]{@{}c@{}}EC-NR\\ ($\langle \Delta I^e_{N_1} \Delta I^e_{N_2} \rangle + \langle \Delta I^h_{N_1} \Delta I^h_{N_2} \rangle$)\end{tabular} & \begin{tabular}[c]{@{}c@{}}CAR-NR\\ ($\langle \Delta I^e_{N_1} \Delta I^h_{N_2} \rangle + \langle \Delta I^h_{N_1} \Delta I^e_{N_2} \rangle$)\end{tabular} \\
\hline
\multirow{3}{*}{\begin{tabular}[c]{@{}c@{}}Non\\Topological\\\end{tabular}} & Gapful & $s$ & $\checkmark$ & $\times$ & $\times$ & $\times$ \\
\cline{2-7}
& \multirow{2}{*}{Nodal} & $p_x \hat{z}$ & $\checkmark$ & $\times$ & $\times$ & $\times$ \\
\cline{3-7}
& & \begin{tabular}[c]{@{}c@{}}$d_{x^2-y^2},$ \\ $d_{xy}$\end{tabular} & $\times$ & $\checkmark$ & $\times$ & $\times$ \\
\hline
\multirow{3}{*}{Topological} & \begin{tabular}[c]{@{}c@{}}Chiral\\(Gapful)\\\end{tabular} & \begin{tabular}[c]{@{}c@{}}$p_x+i p_y,$ \\~$d_{x^2-y^2}+ i d_{xy} $\end{tabular} & $\checkmark$ & $\times$ & $\times$ & $\times$ \\
\cline{2-7}
& \begin{tabular}[c]{@{}c@{}}Helical\\(Gapful)\\\end{tabular} & $p$ & $\checkmark$ & $\checkmark$ & $\checkmark$ & $\checkmark$ \\
\cline{2-7}
& Nodal & \begin{tabular}[c]{@{}c@{}}$p_x\hat{x},$~ \\$p_x \hat{y}$\end{tabular} & $\checkmark$ & $\checkmark$ & $\checkmark$ & $\checkmark$ \\
\hline
\end{tabular}}
\label{pplay trans}
\end{table}

\end{widetext}

Different processes that contribute to shot noise cross-correlations as a function of bias voltages are summarized in Table~\ref{pplay trans} in the low bias voltage range ($-V_1<V_2<V_1$) for transparent limit. All four processes contribute to shot noise cross-correlations for superconductors that do not possess spin rotation symmetry in 2D. In contrast, for superconductors that do not possess spin rotation symmetry, only one process contributes to noise cross-correlations at low bias voltages. We have observed that shot noise cross-correlations for unconventional superconductors do not obey the linearity in bias voltage $V_2$, like $s$ wave.

\section{Experimental Realization \& Conclusion}

\begin{widetext}

\begin{table}[h!]
\centering
\caption{Characteristics of differential non-local conductance and differential shot noise cross-correlations in $N_1$/I/US/I/$N_2$ junction.}
\scalebox{0.84}{\begin{tabular}{|c|c|c|c|c|c|c|c|c|c|c|c|c|}
\hline
\multirow{4}{*}{\begin{tabular}[c]{@{}c@{}}\\Topology\end{tabular}} & \multirow{4}{*}{Type} & \multirow{4}{*}{Pairing} & \multicolumn{2}{c|}{\begin{tabular}[c]{@{}c@{}}Differential\\ non-local conductance\end{tabular}} & \multicolumn{8}{c|}{Differential shot noise cross-correlation} \\
\cline{4-13}
& & & \multicolumn{2}{c|}{\multirow{2}{*}{\begin{tabular}[c]{@{}c@{}}$z=1$, $eV_2 / \Delta=0$\\\end{tabular}}} & \multicolumn{2}{c|}{Symmetric set up} & \multicolumn{2}{c|}{Non-local set up} & \multicolumn{2}{c|}{Symmetric set up} & \multicolumn{2}{c|}{Non-local set up} \\
\cline{6-13}
& & & \multicolumn{2}{c|}{} & \multicolumn{2}{c|}{$z=1$,$eV_1=eV_2$} & \multicolumn{2}{c|}{$z=1$,$eV_2 / \Delta=0$} & \multicolumn{2}{c|}{\begin{tabular}[c]{@{}c@{}}$eV_1 / \Delta=eV_2 / \Delta=0.2$\\\end{tabular}} & \multicolumn{2}{c|}{$eV_1 / \Delta=0.2$,$eV_2 / \Delta=0$} \\
\cline{4-13}
& & & $eV_1/ \Delta=0$ & \begin{tabular}[c]{@{}c@{}}$eV_1= \pm \Delta$\\\end{tabular} & $eV_1/ \Delta=0$ & $eV_1 =\pm \Delta$ & $eV_1/ \Delta=0$ & $eV_1= \pm \Delta$ & Transparent & Tunneling & \begin{tabular}[c]{@{}c@{}}Transparent\\\end{tabular} & \begin{tabular}[c]{@{}c@{}}Tunneling\\\end{tabular} \\
\hline
\multirow{4}{*}{\begin{tabular}[c]{@{}c@{}}Non\\ Topo-\\logical\end{tabular}} & Gapful & $s$ & \begin{tabular}[c]{@{}c@{}}Flat,\\Negative\end{tabular} & Vanishing & \begin{tabular}[c]{@{}c@{}}Flat,\\Positive\end{tabular} & Vanishing & \begin{tabular}[c]{@{}c@{}}Flat,\\Positive\end{tabular} & Vanishing & Positive & Vanishing & Positive & Vanishing \\
\cline{2-13}
& \multirow{3}{*}{Nodal} & $p_x \hat{z}$ & Vanishing & Negative & Vanishing & Positive & Vanishing & Negative & Negative & Vanishing & Negative & Vanishing \\
\cline{3-13}
& & $d_{x^2-y^2}$ & Vanishing & Negative & \begin{tabular}[c]{@{}c@{}}ZBP,\\Positive\end{tabular} & Vanishing & Positive & Vanishing & Vanishing & Positive & Positive & Negative \\
\cline{3-13}
& & $d_{xy}$ & Vanishing & Negative & \begin{tabular}[c]{@{}c@{}}ZBP,\\Positive\end{tabular} & Vanishing & Positive & Vanishing & Vanishing & Positive & Positive & Negative \\
\hline
\multirow{5}{*}{\begin{tabular}[c]{@{}c@{}}Topo-\\logical\end{tabular}} & \multirow{2}{*}{\begin{tabular}[c]{@{}c@{}}Chiral\\(Gapful)\end{tabular}} & $p_x +i p_y$ & \begin{tabular}[c]{@{}c@{}}Flat,\\Negative\end{tabular} & Negative & \begin{tabular}[c]{@{}c@{}}Flat,\\Positive\end{tabular} & Negative & Vanishing & Negative & Positive & Vanishing & Positive & Vanishing \\
\cline{3-13}
& & $d_{x^2-y^2}+ i d_{xy}$ & \begin{tabular}[c]{@{}c@{}}Flat,\\Negative\end{tabular} & Negative & \begin{tabular}[c]{@{}c@{}}Flat,\\Vanishing\end{tabular} & Negative & \begin{tabular}[c]{@{}c@{}}Flat,\\Vanishing\end{tabular} & Negative & Positive & Vanishing & Positive & Vanishing \\
\cline{2-13}
& \begin{tabular}[c]{@{}c@{}}Helical\\(Gapful)\end{tabular} & $p$ & \begin{tabular}[c]{@{}c@{}}ZBD,\\Negative\end{tabular} & Vanishing & \begin{tabular}[c]{@{}c@{}}ZBD,\\Negative\end{tabular} & Vanishing & \begin{tabular}[c]{@{}c@{}}ZBD,\\Negative\end{tabular} & Vanishing & Negative & Positive & Negative & Negative \\
\cline{2-13}
& \multirow{2}{*}{Nodal} & $p_x \hat{x}$ & Vanishing & Negative & \begin{tabular}[c]{@{}c@{}}ZBP,\\Positive\end{tabular} & Vanishing & \begin{tabular}[c]{@{}c@{}}ZBD,\\Negative\end{tabular} & Vanishing & Positive & Positive & Positive & Positive \\
\cline{3-13}
& & $p_x \hat{y}$ & Vanishing & Negative & \begin{tabular}[c]{@{}c@{}}ZBP,\\Positive\end{tabular} & Vanishing & \begin{tabular}[c]{@{}c@{}}ZBD,\\Negative\end{tabular} & Vanishing & Positive & Positive & Positive & Positive \\
\hline
\end{tabular}}
\label{table dsn}
\end{table}

\end{widetext}

Experiments on $NSN$ junctions, similar to that shown in Fig.~\ref{fig:NUSN}, but with $s-wave$ superconductors, are already a decade old. In Ref.~\cite{nsnexpt,nsnexpt1}, positive shot noise cross-correlations were experimentally observed for the first time in a $NSN$ junction with Copper being the metal and Aluminium as a $s-wave$ superconductor. Next, in Ref.~\cite{nsnexpt2}, when Gold replaced Copper as the metal in the $NSN$ junction, similar positive shot noise cross-correlations were again seen. Uniquely, in Ref.~\cite{nsnexpt2}, the effects of an external magnetic field on shot noise cross-correlations were also taken into account. Finally, in a more recent experiment, shot noise auto-correlations were measured in a Metal-High T$_c$ cuprate superconductor junction\cite{nsnexpt3}. Extending these experimental setups to metal-2D unconventional superconductor-metal junctions and measuring shot noise cross-correlations should reveal the signatures of the distinct pairing symmetries. Suggested pairing symmetries for Sr$_2$RuO$_4$ are still in conflict but recent anisotropic strain experiment in Sr$_2$RuO$_4$ suggests pairing as chiral-$p$~\cite{strain}.

In Table~\ref{table dsn}, we summarize the results of our work concerning differential non-local conductance and differential shot noise cross-correlations in both symmetric and non-local setups. Whether we consider symmetric or non-local setup, differential shot noise cross-correlations for helical-$p$ superconductor are always negative in the transparent limit. Further, for the entire range of bias voltages, shot noise is negative for helical-$p$ pairing. It is the unique signature of helical-$p$ pairing. In the transparent and tunnel limits, we summarize the results of shot noise cross-correlations in Table~\ref{table sn}. Irrespective of whether tunnel limit or transparent, HBT correlations for non-topological pairings are always asymmetric to the sign of bias voltage and symmetric for topological pairings.

Our approach using non-local differential conductance, differential shot noise cross-correlations, and shot noise cross-correlations to probe chiral ($p$ and $d$), as well as helical-$p$ and nodal pairing in topological superconductors, will help distinguish helical from chiral and nodal pairing, unlike Knight shift measurement that does not resolve the helical and chiral dichotomy. Our method will give an easy way for experimentalists to distinguish non-topological superconductors from chiral, nodal as well as helical-$p$ superconductors via differential shot noise cross-correlations and shot noise cross-correlations. {We have considered a finite but small value of Rashba spin-orbit coupling $\lambda$ for cases without spin rotation symmetry, i.e., $p_x \hat{x}$, $p_x \hat{y}$ and helical-$p$ which has a minor effect on the magnitude of the results as compared to that for $\lambda=0$.}. In the future, we will extend our study to probe the topological character of superconducting Dirac materials using shot noise cross-correlations.

\begin{widetext}

\begin{table}[h!]
\centering
\caption{Characteristics of shot noise cross-correlations in $N_1$/I/US/I/$N_2$ junction.}
\scalebox{0.8}{\begin{tabular}{|c|c|c|c|c|c|l|c|c|c|l|}
\hline
\multirow{3}{*}{Topoloogy} & \multirow{3}{*}{Type} & \multirow{3}{*}{Pairing} & \multicolumn{8}{c|}{\begin{tabular}[c]{@{}c@{}}Shot noise cross-correlations \\$(eV_1/\Delta=0.2)$\end{tabular}} \\
\cline{4-11}
& & & \multicolumn{4}{c|}{$z=0.1$} & \multicolumn{4}{c|}{$z=3$} \\
\cline{4-11}
& & & $eV_2/ \Delta \rightarrow 0$ & $eV_2= eV_1$ & $eV_2=-eV_1$ & $-eV_1<eV_2<eV_1$ & $eV_2/ \Delta \rightarrow 0$ & $eV_2= eV_1$ & $eV_2=-eV_1$ & $-eV_1<eV_2<eV_1$ \\
\hline
\multirow{3}{*}{\begin{tabular}[c]{@{}c@{}}Non\\ topological\end{tabular}} & Gapful & $s$ & Positive & \begin{tabular}[c]{@{}c@{}}Finite,\\ Positive\end{tabular} & Vanishing & $eV_1+eV_2$ & Vanishing & Positive & Negative & $eV_2$ \\
\cline{2-11}
& \multirow{2}{*}{Nodal} & $p_x \hat{z}$ & \begin{tabular}[c]{@{}c@{}}Flat,\\Positive\end{tabular} & Positive & Vanishing & $(c~eV_2+eV_1)^2$ & \begin{tabular}[c]{@{}c@{}}Flat,\\Positive\end{tabular} & Positive & Negative & $(eV_2+eV_1)^2-c$ \\
\cline{3-11}
& & $d_{xy},d_{x^2-y^2}$ & \begin{tabular}[c]{@{}c@{}}Flat,\\Positive\end{tabular} & Vanishing & Positive & $(c~eV_2-eV_1)^2$ & \begin{tabular}[c]{@{}c@{}}Flat,\\Negative\end{tabular} & Vanishing & Negative & $-(c~eV_2-eV_1)^2$ \\
\hline
\multirow{3}{*}{Topological} & \begin{tabular}[c]{@{}c@{}}Chiral\\ (Gapful)\end{tabular} & \begin{tabular}[c]{@{}c@{}}$p_x +i p_y,$\\$d_{x^2-y^2}+ i d_{xy}$ \end{tabular} & \begin{tabular}[c]{@{}c@{}}ZBP,\\ Positive\end{tabular} & Vanishing & Vanishing & $eV_1-|eV_2|$ & \begin{tabular}[c]{@{}c@{}}ZBP,\\ Positive\end{tabular} & Vanishing & Vanishing & $eV_1-|eV_2|$ \\
\cline{2-11}
& \begin{tabular}[c]{@{}c@{}}Helical\\ (Gapful)\end{tabular} & $p$ & Negative & Vanishing & Vanishing & $(eV_2)^2-(eV_1)^2$ & Vanishing & Vanishing & Vanishing & $-(eV_1|eV_2|-(eV_2)^2)$ \\
\cline{2-11}
& Nodal & $p_x \hat{x}$, $p_x \hat{y}$ & Negative & Vanishing & Vanishing & $(eV_2)^2-(eV_1)^2$ & \begin{tabular}[c]{@{}c@{}}ZBP,\\Positive\end{tabular} & Vanishing & Vanishing & $(-|eV_2|+c~ eV_1)-|(eV_2)^2-(eV_1)^2|$ \\
\hline
\end{tabular}}
\label{table sn}
\end{table}

\end{widetext}

\acknowledgments This work was supported by the grants "Josephson junctions with strained Dirac materials and their application in
quantum information processing" from Science \& Engineering Research Board (SERB), New Delhi, Government of India, under Grant No. $CRG/2019/006258$.

\appendix

\counterwithin{figure}{section}

\section{Differential non-local conductance}
\begin{figure}[h!]
\centering
\includegraphics[scale=1.6]{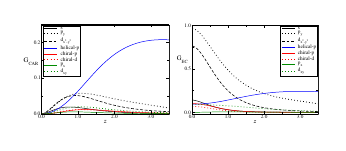}
\caption{Crossed Andreev conductance $G_{CAR}$ (Left) and elastic cotunneling $G_{EC}$ (Right) for $N_1$/I/US/I/$N_2$ junction vs barrier strength ($z$) for US with different pairing symmetries with $k_F R=55$, $R / \xi =2$, $eV_1 / \Delta=0.2$ and $eV_2 / \Delta=0$.}
\label{G}
\end{figure}
In Fig.~\ref{G}, we plot $G_{CAR}$ and $G_{EC}$ vs. barrier strength($z$) for non-local set up with $eV_1 /\Delta=0.2$, $eV_2/ \Delta=0.0$. $G_{CAR}$ for helical-$p$ and non-topological nodal superconductor tends to a finite value in the tunnel limit, while for other cases tends to zero. $G_{EC}$ for helical-$p$ and non-topological nodal superconductors tend to a finite value, but $G_{EC}$ for other cases tend to zero in the tunnel limit ($z \rightarrow$ large). $G_{NL}$ is always dominated by $G_{EC}$.

\begin{figure}[h!]
\includegraphics[scale=0.2]{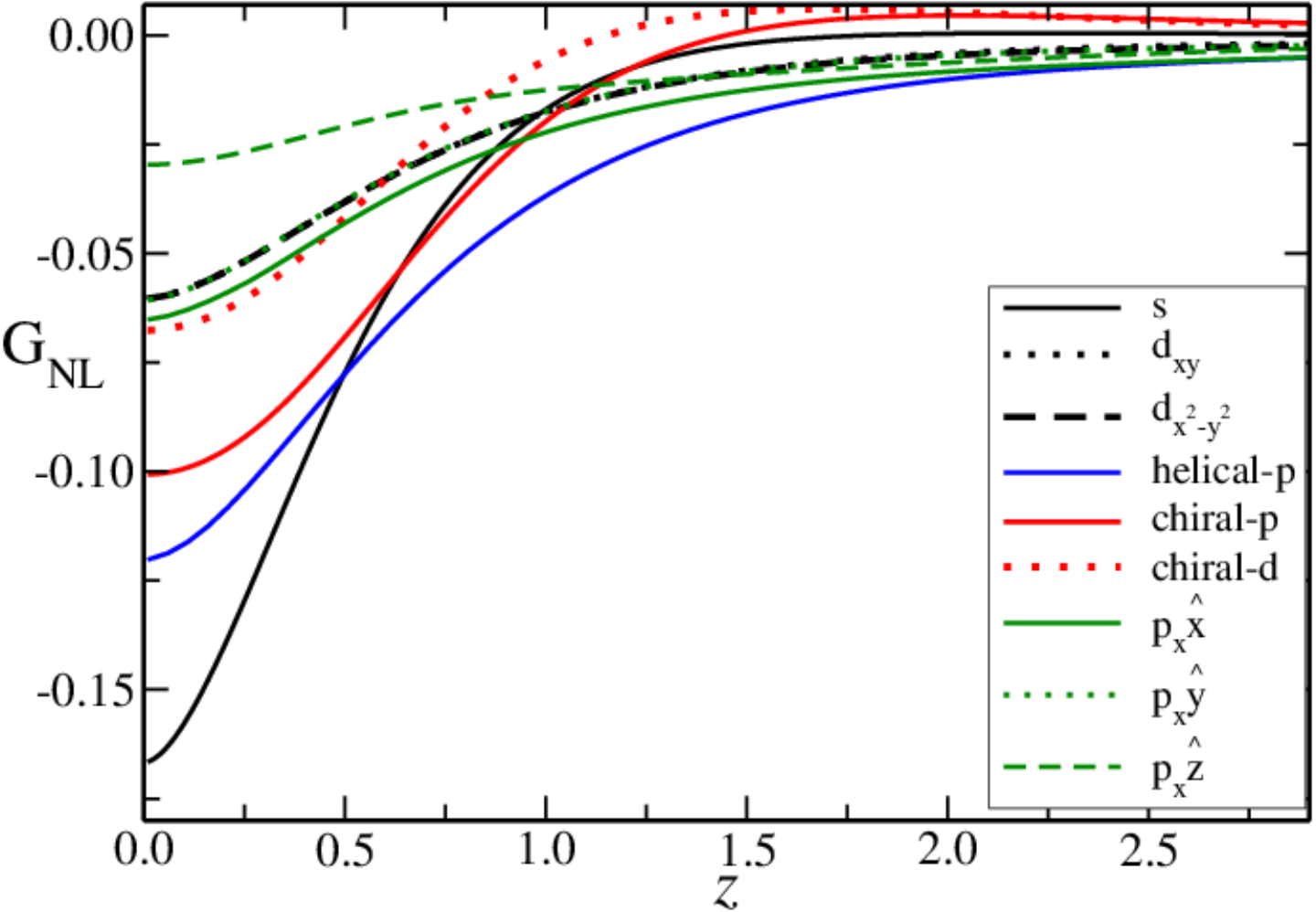}
\caption{$G_{NL}$ for $N_1$/I/US/I/$N_2$ junction vs barrier strength ($z$) for US with different pairing symmetries where $k_F R=55$, $R / \xi =2$, $e$V$_1/ \Delta=0.2$ and $e$V$_2/ \Delta=0$. }
\label{GNLz}
\end{figure}
In Fig.~\ref{GNLz}, we plot $G_{NL}$ vs. $z$ in non-local set up with $eV_1/ \Delta=0.2 $, $eV_2/ \Delta=0.0$. $G_{NL}$ for all pairings tends to zero in the tunnel limit ($z \rightarrow$ large). One does not see any marked difference between topological and non-topological superconductors from the non-local conductance. It is because the electron co-tunneling conductance dominates the crossed Andreev conductance.

\section{Shot noise cross-correlation}
This section expands the shot noise cross-correlations formula regarding scattering amplitudes.
Shot noise cross-correlations as given in Eq.~\ref{eqn:sn} can be expanded as
\begin{eqnarray}
S^{12}&&=\frac{4e^2}{h} \int^{\pi/2}_{-\pi/2} d\theta \frac{\cos{\theta}}{2 \pi} \int dE \{ (s_1+s_2) h_1 + (s_3+s_4) h_2 \nonumber \\
&& + (s_5+s_6) h_3 + (s_7+s_8) h_4 + s_9 h_5 + s_{10} h_6 + s_{11} h_7 + s_{12} h_8 \nonumber \\
&& + s_{13} h_9 + s_{14} h_{10} \},
\label{eqn:snheaviside}
\end{eqnarray}
Fermi function as given in Eq.~2 for electron and hole in contact $i$ are $f_{ie}(E)= \left[ 1 + exp \left( \dfrac{E - |e|V_i}{k_B T} \right) \right]^{-1}$ and $f_{ih}(E)= \left[ 1 + exp\left( \dfrac{E + |e|V_i}{k_B T} \right) \right]^{-1}$, where $k_B$ is Boltzmann constant and $T$ is temperature. At zero temperature, $f_{ie} (E) \rightarrow \Theta(|e|V_i -E)$ and $f_{ih} (E) \rightarrow \Theta(|e|V_i +E)$, where $\Theta$ is Heaviside theta function.

The variables $s_i$ to $i=1,...14$ and Heaviside theta function components $h_i$ to $i=1,...10$, given in Eq.~\ref{eqn:snheaviside} are-

\begin{widetext}
\begin{align}
h_1 &= \Theta (|e|V_1 - E) - 2 \Theta (|e|V_1 - E) \Theta (|e|V_2 - E) + \Theta (|e|V_2 - E), h_2 = \Theta (-|e|V_1 - E) - 2 \Theta (-|e|V_1 - E) \Theta (-|e|V_2 - E) + \Theta (-|e|V_2 - E), \nonumber \\
h_3 &= -\Theta (-|e|V_1 - E) + 2 \Theta (-|e|V_1 - E) \Theta (|e|V_2 - E) - \Theta (|e|V_2 - E), h_4 = -\Theta (|e|V_1 - E) + 2 \Theta (|e|V_1 - E) \Theta (-|e|V_2 - E) - \Theta (-|e|V_2 - E), \nonumber \\
h_5 & = -\Theta (-|e|V_1 - E) + 2 \Theta (-|e|V_1 - E) \Theta (-|e|V_2 - E) - \Theta (-|e|V_2 - E), h_6 = -\Theta (|e|V_1 - E) + 2 \Theta (|e|V_1 - E) \Theta (|e|V_2 - E) - \Theta (|e|V_2 - E), \nonumber \\
h_7 &= \Theta (-|e|V_1 - E) - 2 \Theta (-|e|V_1 - E) \Theta (|e|V_2 - E) + \Theta (|e|V_2 - E), h_8 = \Theta (|e|V_1 - E) - 2 \Theta (|e|V_1 - E) \Theta (-|e|V_2 - E) + \Theta (-|e|V_2 - E), \nonumber \\
h_9 &= \Theta (|e|V_2 - E) - 2 \Theta (|e|V_2 - E) \Theta (-|e|V_2 - E) + \Theta (-|e|V_2 - E), h_{10} = \Theta (|e|V_1 - E) - 2 \Theta (|e|V_1 - E) \Theta (-|e|V_1 - E) + \Theta (-|e|V_1 - E), \nonumber \\
&s_1 = \{ s^{ee}_{12} s^{ee *}_{11} s^{ee}_{21} s^{ee *}_{22} \}, s_2= \{ s^{he}_{21} s^{he *}_{11} s^{he}_{12} s^{he *}_{22} \}, s_3 = \{ s^{hh}_{12} s^{hh *}_{11} s^{hh}_{21} s^{hh *}_{22} \}, s_4 = \{ s^{eh}_{12} s^{eh *}_{11} s^{eh}_{21} s^{eh *}_{22} \}, \nonumber \\
& s_5 = \{ s^{eh}_{21} s^{hh *}_{11} s^{he}_{12} s^{ee *}_{22} \}, s_6= \{s^{ee}_{12} s^{eh *}_{11} s^{hh}_{21} s^{he *}_{22} \}, s_7 = \{ s^{eh}_{12} s^{ee *}_{11} s^{he}_{21} s^{hh *}_{22} \}, s_8 = \{ s^{hh}_{12} s^{eh *}_{22} s^{ee}_{21} s^{he *}_{11} \}, \nonumber \\
& s_9 = \{ s^{eh}_{12} s^{hh *}_{22} s^{hh}_{21} s^{eh *}_{22} + s^{hh}_{12} s^{hh *}_{11} s^{eh}_{21} s^{eh *}_{22} \}, s_{10} = \{ s^{ee}_{12} s^{ee *}_{11} s^{he}_{21} s^{he *}_{22} + s^{ee}_{21} s^{ee *}_{22} s^{he}_{12} s^{he *}_{11} \},\nonumber \\
& s_{11} = \{ s^{ee}_{12} s^{ee *}_{22} s^{eh}_{21} s^{eh *}_{11} + s^{hh}_{21} s^{hh *}_{11} s^{he}_{12} s^{he *}_{22} \}, s_{12} = \{ s^{eh}_{12} s^{ee *}_{11} s^{ee}_{21} s^{eh *}_{22} + s^{hh}_{12} s^{hh *}_{22} s^{he}_{21} s^{he *}_{11} \}, \nonumber \\
& s_{13} = \{ s^{eh}_{11} s^{ee *}_{11} s^{ee}_{21} s^{eh *}_{21} + s^{hh}_{21} s^{hh *}_{11} s^{he}_{11} s^{he *}_{21} - s^{hh}_{11} s^{eh *}_{21} s^{ee}_{21} s^{he *}_{11} - s^{ee}_{11} s^{he *}_{21} s^{hh}_{21} s^{eh *}_{11} \}, \nonumber \\
& s_{14} = \{ s^{ee}_{12} s^{ee *}_{22} s^{eh}_{12} s^{eh *}_{12} + s^{eh}_{12} s^{hh *}_{22} s^{he}_{22} s^{he *}_{12} - s^{eh}_{12} s^{ee *}_{12} s^{he}_{22} s^{hh *}_{22} - s^{hh}_{12} s^{he *}_{12} s^{ee}_{22} s^{eh *}_{22} \}.
\label{eqn:heaviside}
\end{align}
$EC-NR$ process are identified by $s_{1}$, and $s_{3}$, while $CAR-AR$ process are $s_{2}$, and $s_{4}$. $CAR-NR$ process are identified by $s_5$ and $s_7$, while $EC-AR$ process are $s_6$ and $s_8$. Finally, $s_{9}-s_{14}$ are identified as mixed processes that consist of scattering amplitudes of all four processes, i.e., EC, CAR, AR, and NR.
\end{widetext}

\end{document}